\pdfoutput=1

\documentclass[aps,pre,twocolumn,superscriptaddress,showpacs]{revtex4}
\usepackage{graphicx,color}
\usepackage{epstopdf}
\usepackage{amsmath,amssymb, bm}

\usepackage[utf8]{inputenc}
\usepackage{lipsum}

\usepackage{float}
\usepackage[caption=false]{subfig}

\usepackage{fancyhdr}     

\newcommand{\ff}[2]{\frac{#1}{#2}}


\begin{document}

\title{Chaos synchronization by resonance of multiple delay times}
\author{Manuel Jimenez Martin}
\affiliation{Dpto. Fisica Fundamental, Universidad Nacional Educaci\'on a Distancia, 
C/ Senda del Rey 9, 28040 Madrid, Spain}
\author{Otti D'Huys}
\affiliation{Physics Department, Duke University, Box 90305, 120 Science Drive, Durham NC 27708, USA}
\affiliation{Institute of Theoretical Physics, University of W\"urzburg, Am Hubland, 
97074 W\"urzburg, Germany}
\author{Laura Lauerbach}
\affiliation{Institute of Theoretical Physics, University of W\"urzburg, Am Hubland, 
97074 W\"urzburg, Germany}
\author{Elka Korutcheva}
\affiliation{Dpto. Fisica Fundamental, Universidad Nacional Educaci\'on a Distancia, 
C/ Senda del Rey 9, 28040 Madrid, Spain}
\affiliation{G. Nadjakov Inst. Solid State Physics, Bulgarian Academy of Sciences, 1784, Sofia, Bulgaria}
\author{Wolfgang Kinzel}
\affiliation{Institute of Theoretical Physics, University of W\"urzburg, Am Hubland, 
97074 W\"urzburg, Germany}

\date{\today}

\begin{abstract}
Chaos synchronization may arise in networks of nonlinear units with delayed couplings.
We study complete and sublattice synchronization generated by resonance of two large time delays with a specific ratio. 
As it is known for single delay networks, the number of synchronized sublattices is determined by the Greatest 
Common Divisor (GCD) of the network loops lengths. 
We demonstrate analytically the GCD condition in networks of iterated Bernouilli maps with multiple delay times
and complement our analytic results by numerical phase diagrams, providing
parameter regions showing complete and sublattice synchronization by resonance for Tent and Bernouilli maps.
We compare networks with the same GCD with single and multiple delays, and we investigate the sensitivity of 
the correlation to a detuning between the delays in a network of coupled  Stuart-Landau oscillators.
Moreover, the GCD condition also allows to detect time delay resonances leading to 
high correlations in non-synchronizable networks.  Specifically, GCD-induced resonances are observed both in a 
chaotic asymmetric network and in doubly connected rings of delay-coupled noisy linear oscillators.
\end{abstract}

\maketitle

%
\section{Introduction}\label{s:intro}
Chaos synchronization appears in networks of interacting non-linear units \cite{Boccaletti2002, Pikovsky2001}. 
Due to the finite velocity of signal transmission, the couplings might be delayed.
Time delay may generate instability, 
therefore such a network tends to become chaotic \cite{Erneux2009}. On the other side, interaction enforces 
synchronization. 
Under certain circumstances, even when the time delay is much larger than the internal time scale, the units lock to
a common chaotic trajectory without any time shift \cite{Atay2004, Lakshmanan2011}. 
The phenomenon has been demonstrated both numerically and experimentally in small networks 
of nonlinear oscillators \cite{Murphy2009} and coupled semiconductor 
lasers \cite{Mirasso2002,  Fischer06, PeilFischer07, Soriano2013}, and is of natural interest in the fields of neuroscience
\cite{Buzsaki2009, Keane2012} and secure communication \cite{Shore2005}.

For any network of identical non-linear units with a single time delay,
the stability of chaos synchronization is determined by 
the maximum Lyapunov exponent of a single unit with delayed feedback, and the second largest eigenvalue of the adjacency matrix.
Firstly, two chaotic regimes are possible depending on the scaling of the single unit Lyapunov exponent, 
namely strong and weak chaos. Synchronization is only possible in the regime of weak chaos, 
where the exponent is positive and scales inversely with the delay time
\cite{Heiligenthal2011}. Secondly, the stability of the synchronized trajectory is determined by the difference in magnitude 
between the largest and the second largest eigenvalue \cite{Scholl2010}. 
If there is a gap, stable chaos synchronization is possible, otherwise it is ruled out.  
For example, a ring of nonlinear units with unidirectional 
bonds has no eigenvalue gap, hence it cannot synchronize \cite{Heiligenthal2001, buldu2007}. 

More specifically, the number of synchronized 
groups, is determined by the greatest common divisor (GCD) of the network loop lengths \cite{KanterKinzel2011}. 
Complete synchronization is linked with a non-zero eigenvalue gap and is possible 
if $GCD=1$. Additionally, for $GCD=K$, the network shows a pattern of $K$ synchronized groups,
where units belonging to the same group are not connected to each other, only to units from other synchronization groups.
This is called sublattice synchronization \cite{KestlerKanter2007}.

The GCD condition is exact for networks with a single large delay time  and is related to mixing of information 
between the units \cite{KanterCohen2011, KanterKinzel2011}. However, it has been argued that it is also true for networks with multiple large delay times 
with a fixed ratio, as resonances between the delayed signals influence the stability 
of synchronization.  While the problem of 2 coupled units is solved \cite{Zigzag2010},
an analytic proof is still not available for general networks. Nontheless, the 
extended GCD condition has been  
demonstrated in numerical simulations and in experiments on semiconductor lasers: 
in \cite{Rosenbluh2010}, two lasers interacting by transmitting their laser beams 
with a single delay time, become chaotic but cannot synchronize at zero lag. 
When a second, twice as long delay time is added with beam splitters, the two lasers 
could synchronize to a common chaotic intensity.
Hence, adding the second delay time produced synchronization. 
More recently \cite{Nixon2012}, the GCD condition correctly predicted the number of phase synchronization clusters in 
networks of up to 16 coupled lasers.

Our work extends the previous results to small networks with unidirectional bonds and two large delay times.
Our main contributions are two. Firstly, complete and sublattice synchronization triggered by time delay resonance,
as predicted by the GCD condition, are demonstrated analytically in networks of iterated maps.
Secondly, the scope of the GCD condition is investigated beyond chaos synchronization
showing that time delay resonances also trigger high correlations among units in
non-synchronizable networks. 
GCD-induced high correlations are found in two cases, in an asymmetrical chaotic network 
and in doubly connected rings of noisy linear oscillators. 

The paper is structured as follows.
In Section \ref{s:formalism} we generalize the formalism of Master Stability Function \cite{Pecora1998, Kinzel2009}
to  networks with double time delay. 
We provide analytic results for doubly connected rings of Bernouilli maps. 
Complete and sublattice synchronization are explained by means of the master stability function symmetries and
we give special time delay ratios for which complete synchronization is not possible.
In Section \ref{s:phasemaps} we study complete and sublattice synchronization induced by different 
time delay resonances in doubly connected rings of Tent and Bernoulli maps.
We also discuss sensitivity to detuning and compare with equivalent single-delay networks with virtual units.
Section \ref{s:corrs} investigates the validity of the GCD argument in non-synchronizable networks.
We present an asymmetric chaotic network showing high correlations for $GCD=1$. 
We also study doubly connected rings
of noisy linear oscillators, for which we find GCD-induced correlation peaks.
Finally, the results are summarized in Section \ref{s:summ}.
%
%
\section{Coupled chaotic maps with double delays }\label{s:formalism} 

\subsection{Master stability function}
Generally, a network of iterated maps with two time delays can be modeled as follows:
\begin{multline}
u_{t}^{i}=(1-\epsilon)f(u_{t-1}^{i})+\\
\epsilon\sum_{j=1}^{N}\left[(1-\kappa)
G_{ij}^{(1)}f(u_{t-\tau_{1}}^{j})+\kappa
G_{ij}^{(2)}f(u_{t-\tau_{2}}^{j})\right],
\label{eq:MSF1}
\end{multline}
where $f(x):[0,1]\to[0,1]$ is a chaotic map, $\tau_{2}>\tau_{1}$ are the coupling delays and $\epsilon$ and $\kappa$
are coupling strengths ranging between $0$ and $1$. 
The adjacency matrices $G^{(1)}$ and $G^{(2)}$ represent edges with time delays $\tau_1$ and 
$\tau_2$ respectively. Both have unit row sum $\sum_{j}G_{ij}^{(l)}=1$,
ensuring that any trajectory belonging to the synchronization manifold (SM), $u_t^i=s_t$, 
is a solution.
To calculate the stability of the SM, we can study the evolution of a small perturbation around it $\vec{u}_t=s_t+\vec{\delta}_t$.
If the matrices $G^{(1)}$ and $G^{(2)}$ commute, there exists a common base of eigenvectors  $\vec{\omega}_{n}$
with respective eigenvalues $\gamma_n^{(1)}$ and $\gamma_n^{(2)}$, 
and one can decompose the small perturbation into its eigenmodes $\vec{\delta}_t=\sum_n\xi_{n,t}\vec{\omega}_{n}$.
The linear stability of the synchronized state $s_t$, is then determined 
by the evolution of the amplitudes $\xi_{n,t}$ of each mode 
\begin{eqnarray}
\xi_{n,t} & = & (1-\epsilon)f'(s_{t-1})\xi_{n,t-1} + \epsilon (1-\kappa)\gamma_n^{(1)} f'(s_{t-\tau_1})\xi_{n,t-\tau_{1}} \nonumber \\ 
& & +\epsilon\kappa\,\gamma_n^{(2)}f'(s_{t-\tau_2})\xi_{n,t-\tau_{2}}\,.
\label{eq:linearization}
\end{eqnarray}
A generalized Master Stability Function(MSF) is then calculated as 
\begin{equation}
\lambda\left(\gamma_n^{(1)},\gamma_n^{(2)}\right)=\displaystyle\lim_{t\rightarrow\infty}\frac{1}{t}\ln\frac{|\xi_{n,t}|}{|\xi_{n,0}|}.
\end{equation}
The unit row sum guarantees a common eigenvector $\vec{\omega}_0=\left[1,1,...,1\right]$ with eigenvalues
$\gamma_0^{(1)}=\gamma_0^{(2)}=1$. This mode is parallel to the SM and preserves
synchronization. Every other mode $n>0$ is perpendicular to the SM.
If the Lyapunov exponent along the parallel mode is positive, $\lambda(1,1)>0$, the synchronized trajectory
is chaotic. Aditionally, along the transverse modes $\vec{\omega}_{n>0}$, the MSF should be negative, 
$\lambda(\gamma_n^{(1)},\gamma_n^{(2)})<0$ for all $n>0$, to guarantee the stability of the synchronized state.

In this manuscript we consider networks of two different kind of maps. Bernouilli maps, modeled by 
$$f(x)=ax\,\mod\,1\,,$$
are chaotic for $a>\,1$. Since the derivative $f'(x_t)=a$ is constant, their MSF can be calculated analytically.
Moreover, the analytic results are known to reproduce qualitatively several features of 
more complex chaotic delay systems \cite{EnglertKanter2011}. 
For some properties, however, the fluctuations of the derivative play a role \cite{Jungling2015}.
Therefore,  we compare our results to Tent maps, modeled by 
\[f(x)=
\left\{
   \begin{array}{lll}
   \frac{x}{b} & \textnormal{if} & 0 \leq x < b \\
   & & \\
   \frac{1-x}{1-b} & \textnormal{if} & 1 \geq x \geq b\\
  \end{array}\,.
\right.\]
For Bernouilli maps, Eq. \eqref{eq:linearization} has constant coefficients and we can assume
an exponentially evolving perturbation $\xi_{n,t}=\xi_{n,0} z^{t}$. 
Then, we find the characteristic polynomial 
\begin{equation}
1=a(1-\epsilon)z^{-1} + a\epsilon\left[(1-\kappa)\gamma_n^{(1)}z^{-\tau_1}+\kappa \gamma_n^{(2)}z^{-\tau_2}\right]
\label{eq:YPoly}\,
\end{equation}
which has $\tau_2$ complex roots $z_{r}$, with $r=1,… \tau_2$ for each set of eigenvalues 
$\gamma^{(1)}_n,\gamma^{(2)}_n$. The spectrum of Lyapunov exponents along the eigenmode $\vec{\omega}_n$ is then 
given by $\left\{ \lambda_r\right\} = \left\{\ln|z_{r}|\right\}$. A perturbation mode is stable if all the roots $z_{r}$ lie inside the 
unit circle. 
If $a(1-\epsilon)>1$ holds, one immediately finds an unstable solution $z\simeq a(1-\epsilon)$ irrespective 
of $\gamma_n$. This corresponds to the strong chaos regime
where the two delay terms vanish in the limit $\tau_1\,,\tau_2\to\infty$ \cite{Heiligenthal2011}.
The condition for weak chaos is thus given by
\begin{equation}
 \epsilon>\frac{a-1}{a}.
 \label{eq:StrongChaos}
\end{equation}
Assuming both delays to be large, we write $\tau_1=p\tau$ and $\tau_2=q\tau$ and consider the 
limit $\tau \to \infty$. We introduce a weak chaos ansatz, $z_r =\exp[i\phi_r+\frac{l_r}{\tau}]$ and define 
$Y_r=\exp[-i\phi_r\tau - l_r]=z_r^{-\tau}$. In the limit of large delay for weak chaos, we obtain 
$\exp(-l_r/\tau) \to 1$. Then, for each eigenmode $n$, Eq. \eqref{eq:YPoly} becomes
\begin{equation}
1=a(1-\epsilon)e^{-i\phi_r} + a\epsilon\left[(1-\kappa)\gamma_r^{(1)}Y_r^p+\kappa \gamma_r^{(2)}Y_r^q \right]\,.
\label{eq:pseudospec}
\end{equation}
Since for $\tau \rightarrow \infty$, the phases of the multipliers $\phi_r$  
are uniformly distributed between $0$ and $2\pi$, we can solve for $Y(\phi)$, to obtain the so-called 
pseudo-continuous spectrum \cite{Yanchuk2009}, on which the multipliers are densely located: $\l(\phi)=-\ln|Y(\phi)|$. 
The generalized master stability function is then given by 
\begin{equation}
 \lambda(\gamma^{(1)},\gamma^{(2)})= \frac{1}{\tau}\max_{\phi}(l(\phi)).
 \label{eq:MSF}
\end{equation}
Eq. \eqref{eq:pseudospec} and its symmetries are our main tools to study the network synchronization 
properties.
\subsection{Directed rings with multiple delays}
\begin{figure}[b!]
    \subfloat[\label{fig:ringmaps}]{
    \includegraphics[width=0.48\linewidth]{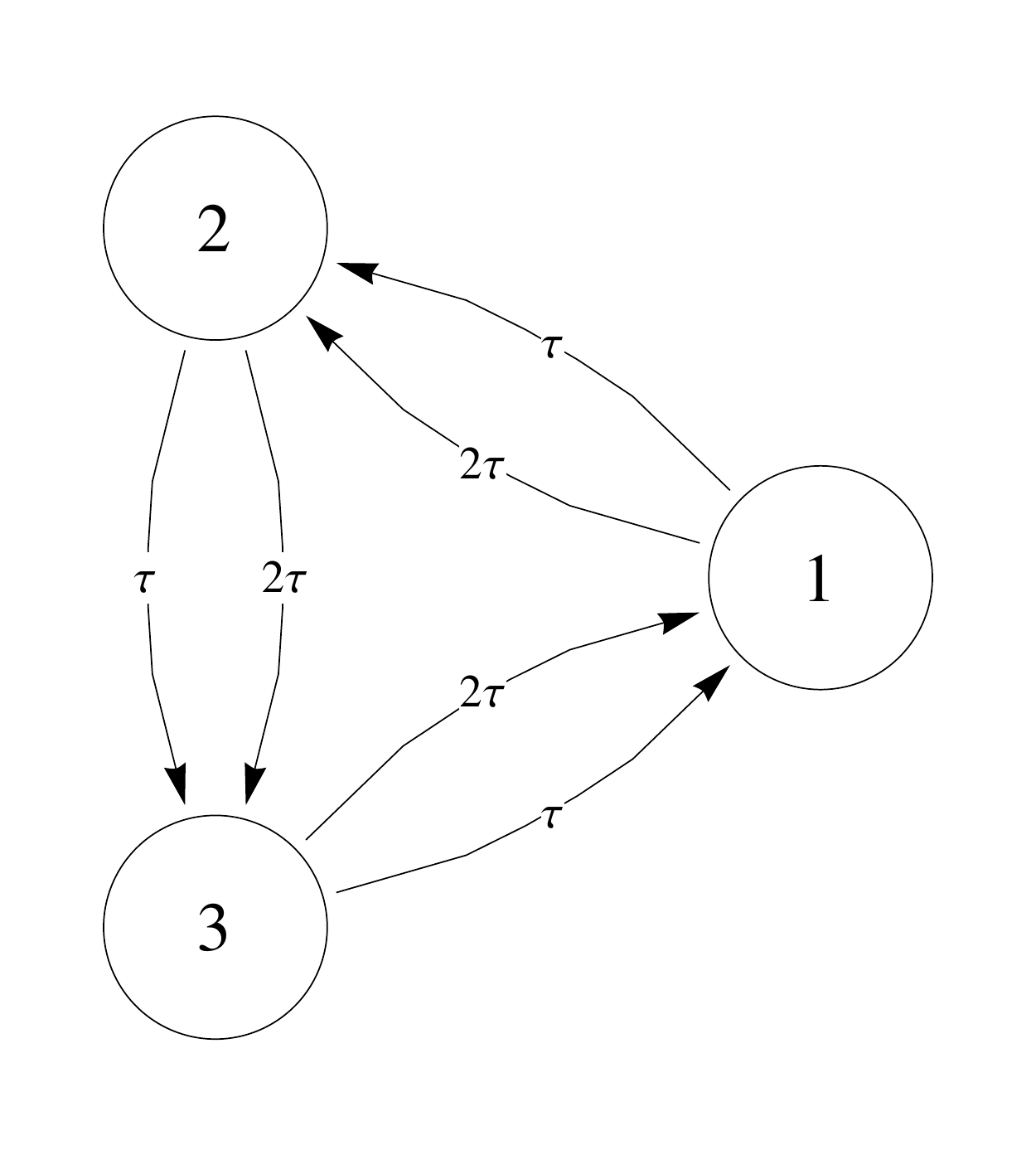}
    }
    \subfloat[\label{fig:ringmaps1}]{
    \includegraphics[width=0.48\linewidth]{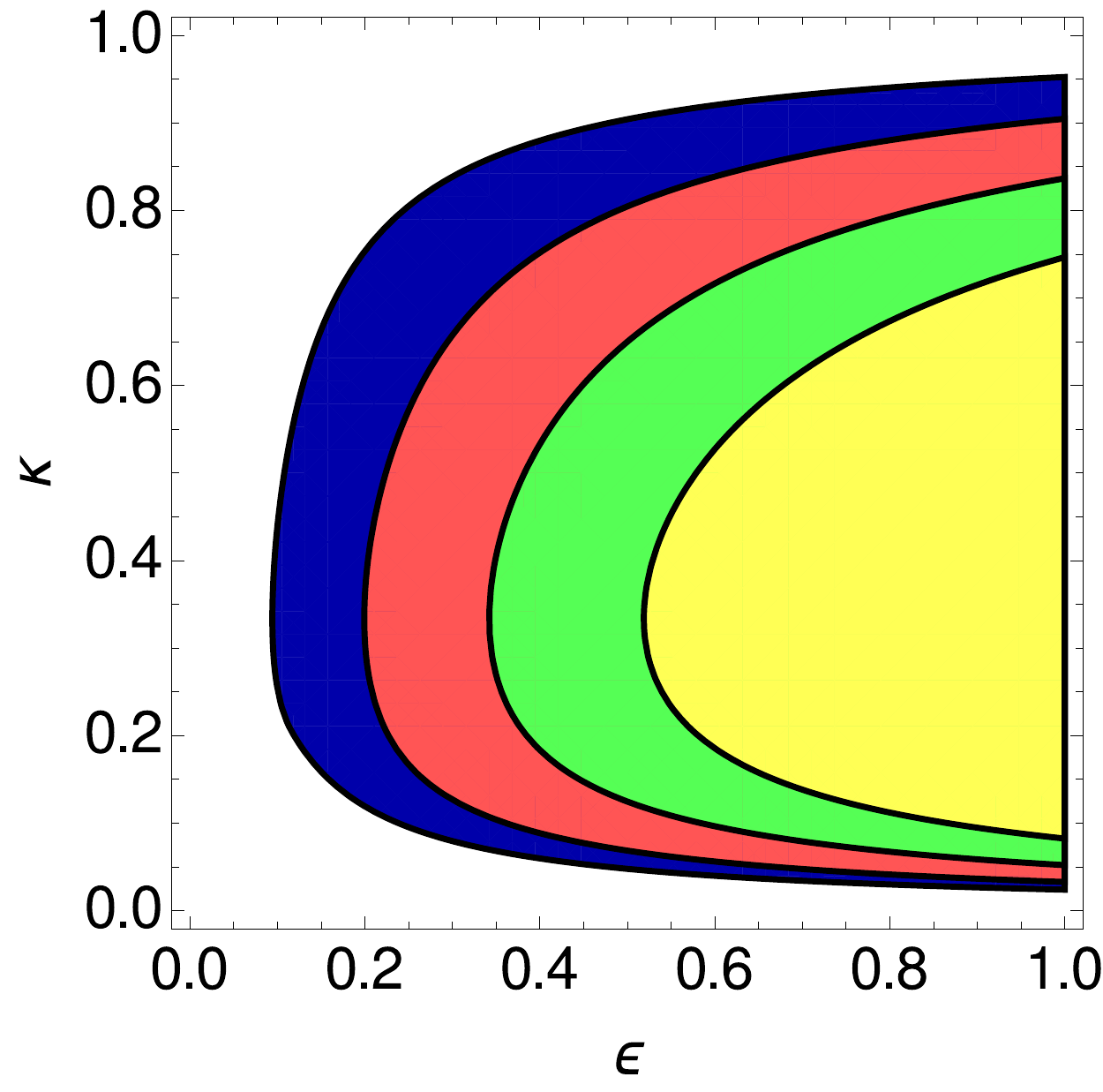}
    }
    \\
    \subfloat[\label{fig:ringmaps2}]{
    \includegraphics[width=0.48\linewidth]{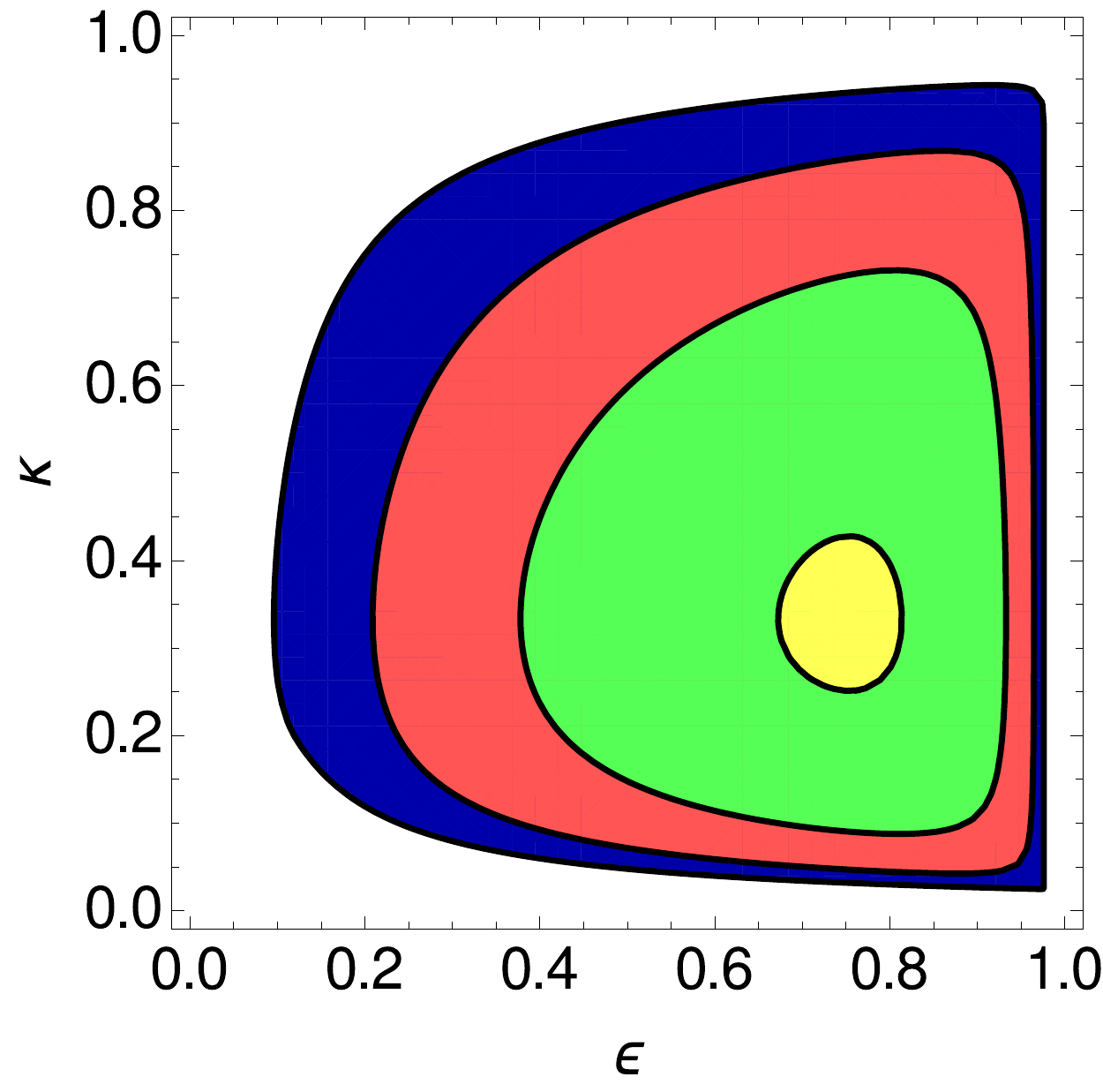} 
    }
    \subfloat[\label{fig:ringmapstents}]{
    \includegraphics[width=0.48\linewidth]{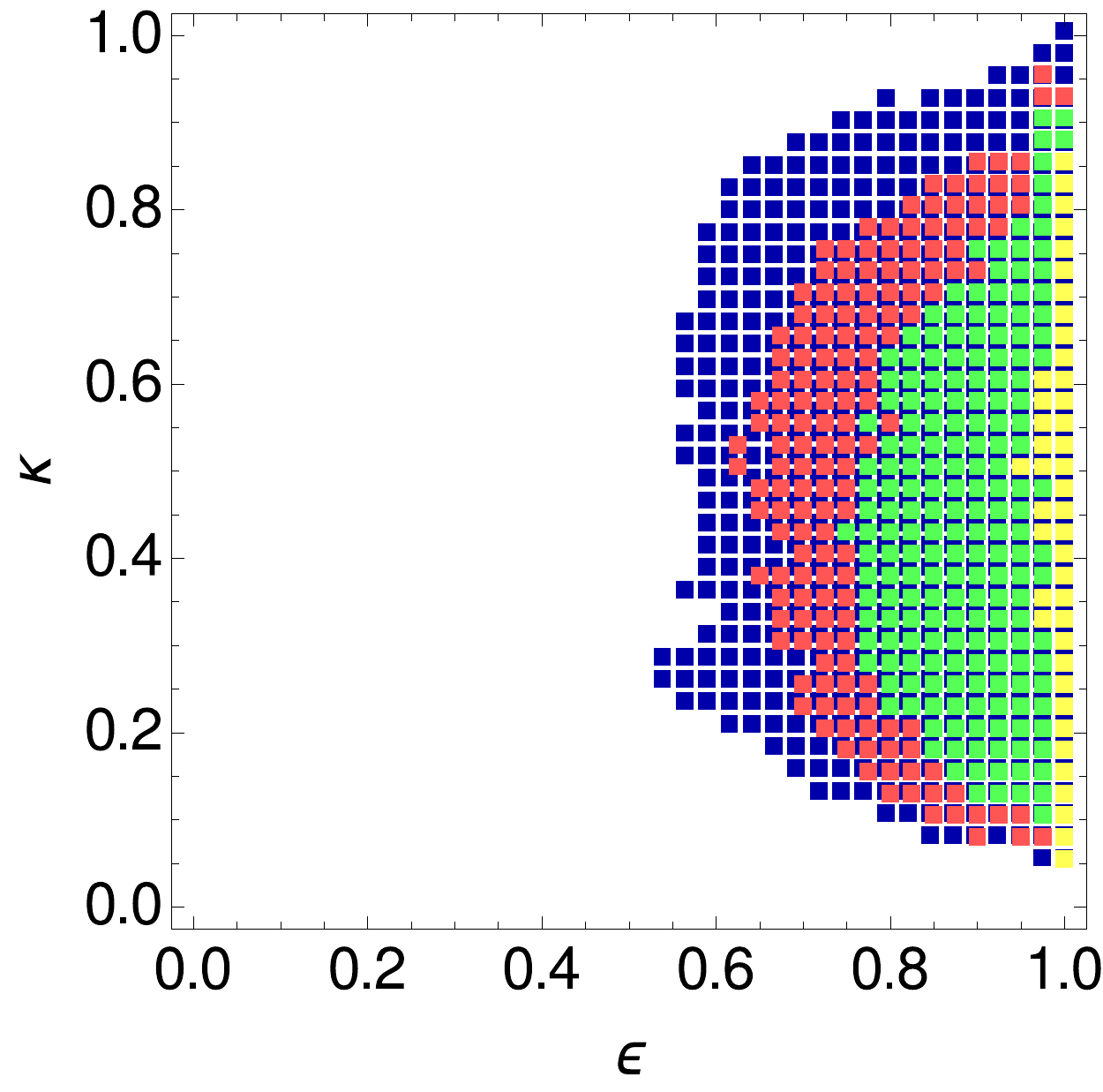} 
    }
\caption{(color online). 
\protect\subref{fig:ringmaps} 
Sketch of a 3 unit directed ring with double time delay.
\protect\subref{fig:ringmaps1} 
Complete synchronization regions for directed rings of Bernouilli maps with  $\tau_2=2\tau_1$ and $a=1.05$.
Synchronization regions are nested 'tongues' of smaller area for increasing number of ring units:
 2 (blue), 3 (red), 4 (green) and 5 (yellow). The color code and nested structure
is common to all subsequent subfigures. The maps are obtained by solving Eq. \eqref{eq:YPoly} 
numerically for every perpendicular mode $n>0$ and then intersecting the stable regions.
\protect\subref{fig:ringmaps2}  
Complete synchronization maps for the same networks after a detuning $(\tau_1,\tau_2) \to (\tau_1-1,\tau_2-1)$. 
\protect\subref{fig:ringmapstents}  
Equivalent complete synchronization regions for rings of Tent maps with equal local Lyapunov exponent:
$b = 0.008458$, $\lambda_L=\log{1.05}$.
Squares indicate completely synchronized trajectory after a small perturbation of
magnitude $10^{-3}$ at the SM and $40000$ map iterations. 
}
\label{fig:RingsBernouilli}
\end{figure}

We will now apply the previous formalism to study directed rings with two time delays (Fig. \ref{fig:ringmaps}).
This is a paradigmatical system that cannot synchronize with a single time-delay, but shows synchronization by time-delay
resonance. The simplest possible ring, consisting on two mutually coupled units $N=2$, was addressed in \cite{Zigzag2010}.
There, it was shown that synchronization with multiple time delays was possible under certain conditions for time delays values. 
It is immediate to check that the provided $\tau_1$, $\tau_2$ values leading to zero lag synchronization 
corresponded to network configurations with $GCD=1$, thus being coherent with our results. 
In this section we generalize to the case of rings of $N$ units with two time delays and no self-feedback. 
We will study the symmetries of the MSF in order to re-derive the time-delay ratios allowing either complete or sublattice 
synchronization as predicted by the GCD condition.

The polynomial \eqref{eq:pseudospec} allows us to study synchronization stability analytically.
$N$-unit doubly coupled unidirectional rings translate to identical adjacency matrices $G^{(1)}=G^{(2)}$ with eigenvalues 
$\gamma_n^{(1)}=\gamma_n^{(2)}\equiv \gamma_n=e^{i\frac{2\pi n}{N}}$, where $0\leq n<N$. 
Thus, the pseudo-continuous spectrum Eq. \eqref{eq:pseudospec} becomes
\begin{equation}
1=a(1-\epsilon)e^{-i\phi} + \gamma_n a\epsilon\left[(1-\kappa)Y^p+\kappa Y^q \right]
\label{eq:pseudospecring}\,.
\end{equation}
Considering the case of single delay, $\kappa=0$, we can easily solve Eq. \eqref{eq:pseudospecring}, 
and find a MSF
\begin{equation}
  \lambda(\gamma)=-\frac{1}{\tau}\ln\left|\frac{1-a(1-\epsilon)}{\gamma_n a \epsilon}\right|\,.
  \label{eq:Single}
\end{equation}
For a single delay, the stability of a perturbation mode only depends on the magnitude of its 
corresponding eigenvalue 
$|\gamma_n|$, i.e. the master stability function is spherically symmetric in the complex plane
with respect to $\gamma_n$. Consequently, the stability of all the eigenmodes, transverse and parallel, 
is the same. Since $\lambda(\gamma_0)=\lambda(1)>0$ for chaotic dynamics, both complete and sublattice 
synchronization are unstable in unidirectional rings of any size with a single delay.

If we consider networks with two time delays, i.e. $\kappa\neq 0$, this spherical symmetry 
can be broken, and the transversal modes can be stabilized, depending on the ratio of the delays. 
However, depending on the $GCD$ the MSF, $\lambda(\gamma)$, still can have some symmetry. 
It is straightforward to check that the GCD of doubly connected rings is $GCD(Np,(N-1)p+q)=GCD(Np,q-p)$.
Consider $p$ and $q$ relatively prime and $q-p=K$.
If the number of elements in the ring $N$ is a multiple of $K$, we find $GCD(Np, p-q)=K$.
Then, the spectrum contains $K$ eigenvalues of the form 
$\hat{\gamma}_r\equiv e^{\ff{i2\pi r}{K}}$, with $r=0,\dots,K-1$,
where the mode $r=0$ is the parallel mode.
It can be shown that the master stability function $\lambda(\gamma)$ is invariant under 
a transformation $\gamma\rightarrow e^{\frac{i2\pi r}{K}}\gamma$. 
Hence all the modes $\hat{\gamma}_r$ have the same Lyapunov exponent, $\lambda(\hat{\gamma}_r)=\lambda(1)$.
Since the parallel mode is unstable, all of them are unstable as well.

To demonstrate this point, we can write $p=r+lK$ and $q=r+(l+1)K$, for some integers $l$  and $0<r<K$. 
Since $p$ and $q$ are relatively prime, we find that $r$ and $K$ are relatively prime as well. 
Then, the pseudo-continuous spectrum Eq. \eqref{eq:pseudospecring} along the eigenmode with eigenvalue is
$\hat{\gamma}_r=e^{\frac{i2\pi r}{K}}$.
\begin{eqnarray}
  1 & = & a(1-\epsilon)e^{-i\phi} + a\epsilon e^{\frac{i 2\pi r}{K}}\left[ (1-\kappa)Y^p+\kappa Y^q \right] \Leftrightarrow\\
  1 & = & a(1-\epsilon)e^{-i\phi} + a\epsilon\left[(1-\kappa)(e^{\frac{i2\pi}{K}}Y)^p+\kappa (e^{\frac{i2\pi}{K}}Y)^q \right] \nonumber \,.
\end{eqnarray}
Since the MSF $\lambda(\gamma)$ only depends on the magnitude $|Y(\phi)|$, 
we conclude that $\lambda(1)=\lambda(\hat{\gamma}_r)$. The corresponding eigenvectors $\vec{\omega}_r$
have $K$ distinct entries
$\vec{\omega}_r=(\gamma_r,\gamma_r^2,\hdots,\gamma_r^N)$, with a phase difference of $\ff{2\pi r}{K}$.
These unstable modes give thus rise to $K$ different sublattices corresponding to the $K$
distinct entries of $\vec{\omega_r}$. 
%

Moreover, it is possible to rule out beforehand some time delay ratios that do not allow complete synchronization. 
The limit to stability is given by $|z|=1$ for the roots of the characteristic polynomial Eq. \eqref{eq:YPoly}. 
Hence $z=e^{i\psi}$, with $\psi$ uniformly distributed along the unit circle if $\tau$ is 
sufficiently large. Choosing $\psi=\frac{\theta}{\tau}$, we get
\begin{equation}
a\epsilon\kappa\gamma_n  e^{-iq\theta}+
a\epsilon(1-\kappa)\gamma_n e^{-ip\theta}+
a(1-\epsilon) =1\,,
\label{eq:z_stability}
\end{equation}
where we have considered $z^{-1}=e^{i\ff{\theta}{\tau}}\approx1$. For the parallel mode, this reduces to
\begin{equation}
  a\epsilon\kappa e^{-iq\theta}+
  a\epsilon(1-\kappa)e^{-ip\theta}+
  (1-\epsilon)a=1\,.
  \label{eq:z_paralel}
\end{equation}
But Eq. \eqref{eq:z_paralel} also holds for a perpendicular mode $n \neq 0$ whenever both
\begin{equation}
 \gamma_n e^{-iq\theta}=e^{i2\pi l}\qquad\text{ and }\qquad\gamma_n e^{-ip\theta}=e^{i 2\pi m}\,,
 \label{eq:forbidden}
\end{equation}
hold at the same time. By substituting $\gamma_n=e^{i2\pi n/N}$ and taking the quotient 
of the phases of Eqs. \eqref{eq:forbidden}, we find that  mode $n$ will be unstable
for a delay ratio
\begin{equation}
  \frac{\tau_1}{\tau_2}=\frac{p}{q}=\frac{n+lN}{n+mN} \,,
  \label{eq:forbidden_ratio}
\end{equation}
where $l$, $m$ are integers and $l\neq m$. 
%
Once an unstable mode is found, the periodicity of its eigenvector determines
the number of synchronized sublattices. 

We proved this result analytically for Bernouilli maps, however this is a rather general phenomenon. 
The symmetry arguments of the master stability function and Eq. \eqref{eq:YPoly} apply also to steady states and periodic orbits in general
since the derivatives along the trajectory are also constant or periodic in this case. 
It can be argued that these symmetry arguments can be extended to chaotic attractors, 
as these consist of unstable periodic orbits \cite{Scholl2010},
but rigorous analytic proofs for chaotic systems other than Bernouilli maps are difficult. 
We provide numerical evidence for the generality of our analytic results in Section  \ref{s:phasemaps}.

%
\section{Complete and sublattice synchronization in directed rings}\label{s:phasemaps}
In this Section, we demonstrate both complete and sublattice synchronization in rings with  double delay, 
for the cases $p/q=1/2$ and $p/q=1/3$, respectively. 
Moreover, we compare the synchronization regions with those of equivalent single delay networks 
with the same loop lengths. 
We investigate the sensitivity to a small detuning of the two delays as well.

\subsection{Complete synchronization in a directed ring with two delays}\label{s:completesync}
If we choose $p/q=1/2$, a ring of $N$ nonlinear units contains loops of all lengths $(N+j)\tau$, 
with $0\leq j\leq N$. Thus, the GCD is always equal to one. Eq. \eqref{eq:pseudospecring} is  
a second degree polynomial and can be solved for each mode $\gamma_n$. 
The complete synchronization region in $\epsilon$-$\kappa$ space is then the intersection of all the 
transverse modes' stability regions.

\begin{figure}[t!]
  \minipage[t]{0.45\columnwidth}
   \includegraphics[width=\linewidth]{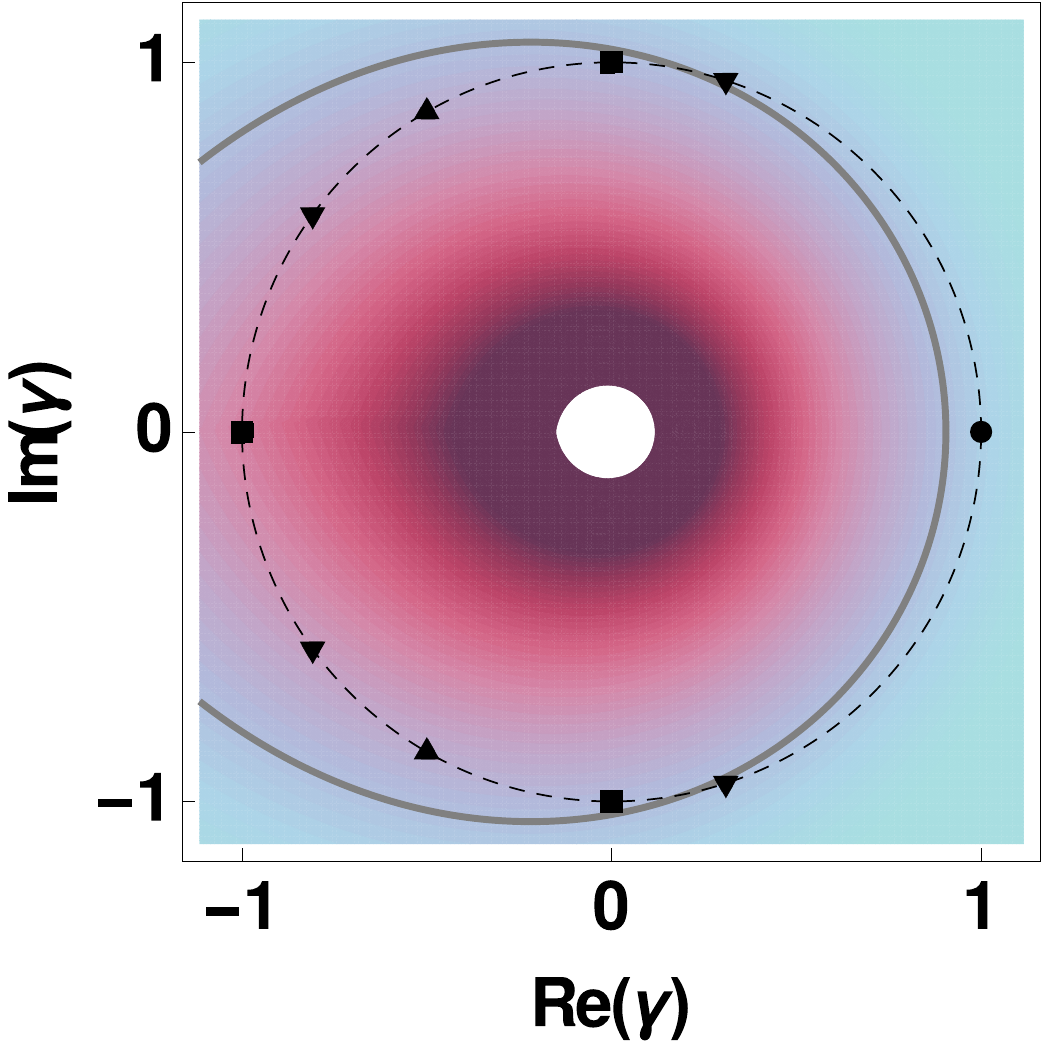}
  \label{fig:MSFa}
  \endminipage
  \minipage[t]{0.45\columnwidth}
   \includegraphics[width=\linewidth]{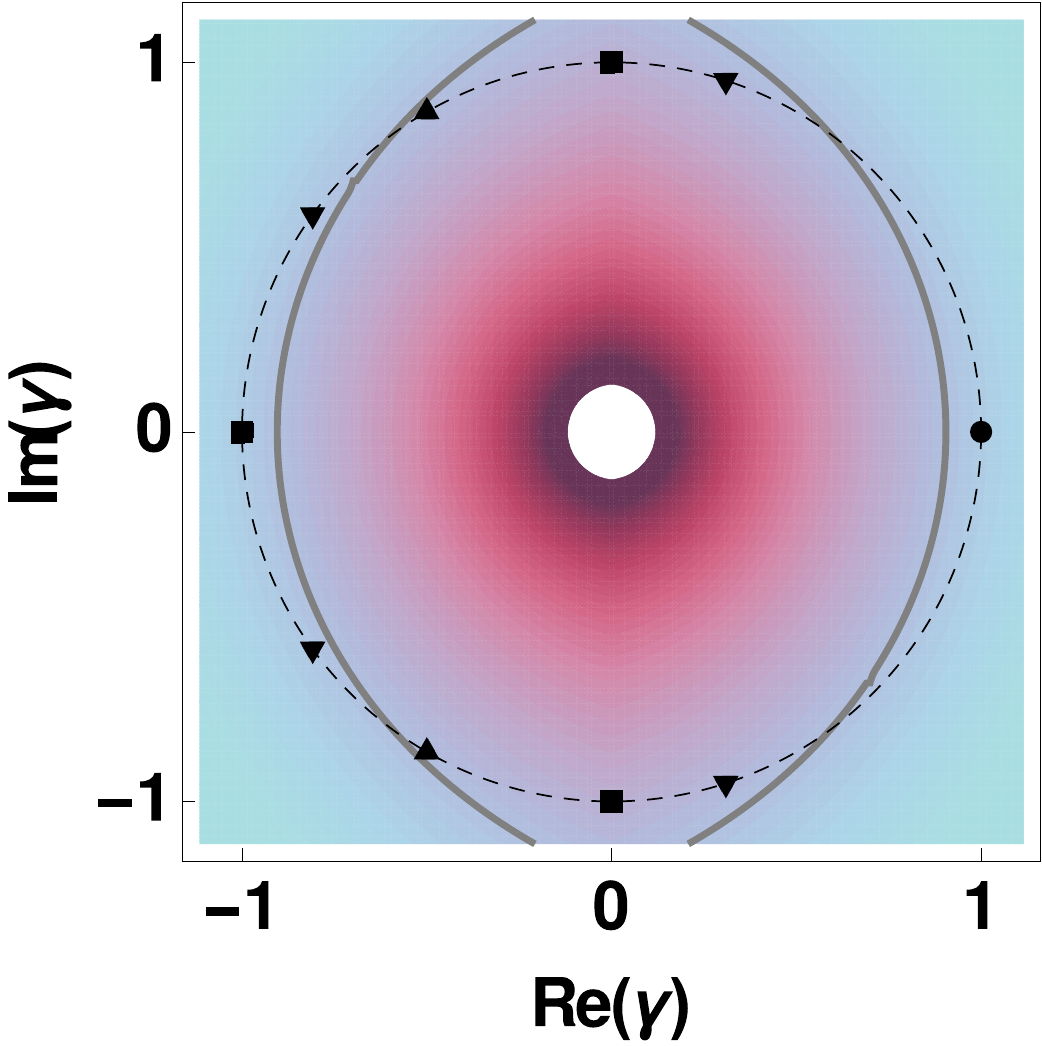}
   \label{fig:MSFb}
  \endminipage
  \hfill
  \minipage[t]{0.082\columnwidth}
  \includegraphics[width=\linewidth]{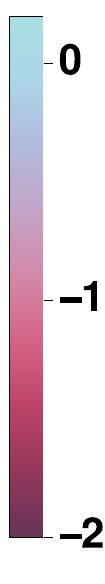}
  \vfill
  \endminipage
  \caption{(color online). Master stability function $\lambda(\gamma,\gamma)$ for a double delayed network 
  as obtained by solving Eq. \eqref{eq:YPoly}, with 
  $2\tau_1=\tau_2$ (left) and $3\tau_1=\tau_2$ (right). 
  Parameters are $\kappa=\epsilon=0.5$ and $a=1.1$. 
  The solid line marks the contour $\lambda=0$. Thus, the eigenmodes whose corresponding eigenvalue 
  lies inside the contour are stable and those staying outside are unstable.
  The eigenvalues $\gamma_n=\exp(2\pi i n/N)$ are  
  indicated for $N=3$ (triangles), $N=4$ (squares) and $N=5$ (inverted triangles). The eigenvalue $\gamma_0=1$ 
  along the synchronization manifold is indicated with a circle.}\label{fig:MSF}
\end{figure}

We show the master stability function $\lambda(\gamma)$ for a ring of Bernouilli maps in Fig. \ref{fig:MSF}. 
The spherical symmetry is clearly broken; the closer the phase of the eigenvalues $\gamma_n$ to $\pi$, 
the smaller the corresponding Lyapunov exponent $\lambda(\gamma_n)$. 
Consequently, the stable parameter region for $\gamma_n=\pi$ is largest. 
It is hence easiest to stabilize zero-lag synchronization for only two coupled elements, 
where this is the only transverse eigenvalue. For our choice of parameters, zero lag synchronization is stable 
for $N=2$, as $\lambda(-1)<0$, as indicated by the square. 
Also for $N=3$, we find $\lambda(e^{2\pi i/3})=\lambda(e^{-2\pi i/3})<0$ in the stable region of 
the MSF, both eigenvalues are indicated by triangles. 
For $N=4$ the MSF is unstable for the modes  $\gamma=\pm i$, 
and for $N=5$ we find unstable transverse modes for $\gamma=e^{\pm 2\pi i/5}$. 
Zero lag synchronization is hence unstable in both cases for the chosen parameters.

For a different parameter choice, the phase maps showing the parameter region for which zero lag 
synchronization is stable in rings of  $N$  Bernouilli maps, are depicted on Fig. \ref{fig:ringmaps1}. The  size of the 
synchronization regions shrinks with increasing number of units, in agreement with the 
shape of the master stability function. The analytic phase diagrams have been confirmed by numerical 
iterations of the chaotic network. 

In order to demonstrate the GCD condition for other chaotic units, for which analytic results are not 
available, we simulated analogous networks of Tent maps.
In order to compare both dynamics, we picked the Tent and Bernouilli maps parameters, $a$ and $b$,
laying the same Lyapunov exponent for the single isolated maps without delay,
$\lambda_L=\log{a}=\log{b^b (1-b)^{1-b}}$.
The resulting synchronization regions are shown in Fig. \ref{fig:ringmapstents}. 
We find a similar structure of nested smaller 
regions for increasing number of units, but due to the fluctuations of the derivative of the map, 
the corresponding coefficients of Eq. \eqref{eq:pseudospec} become time dependent, 
shrinking the regions of stable chaos synchronization.
\subsection{Sublattice synchronization}\label{s:sublatticesync}
\begin{figure}[h!]
    \subfloat[\label{fig:4ring}]{
    \includegraphics[width=0.48\linewidth]{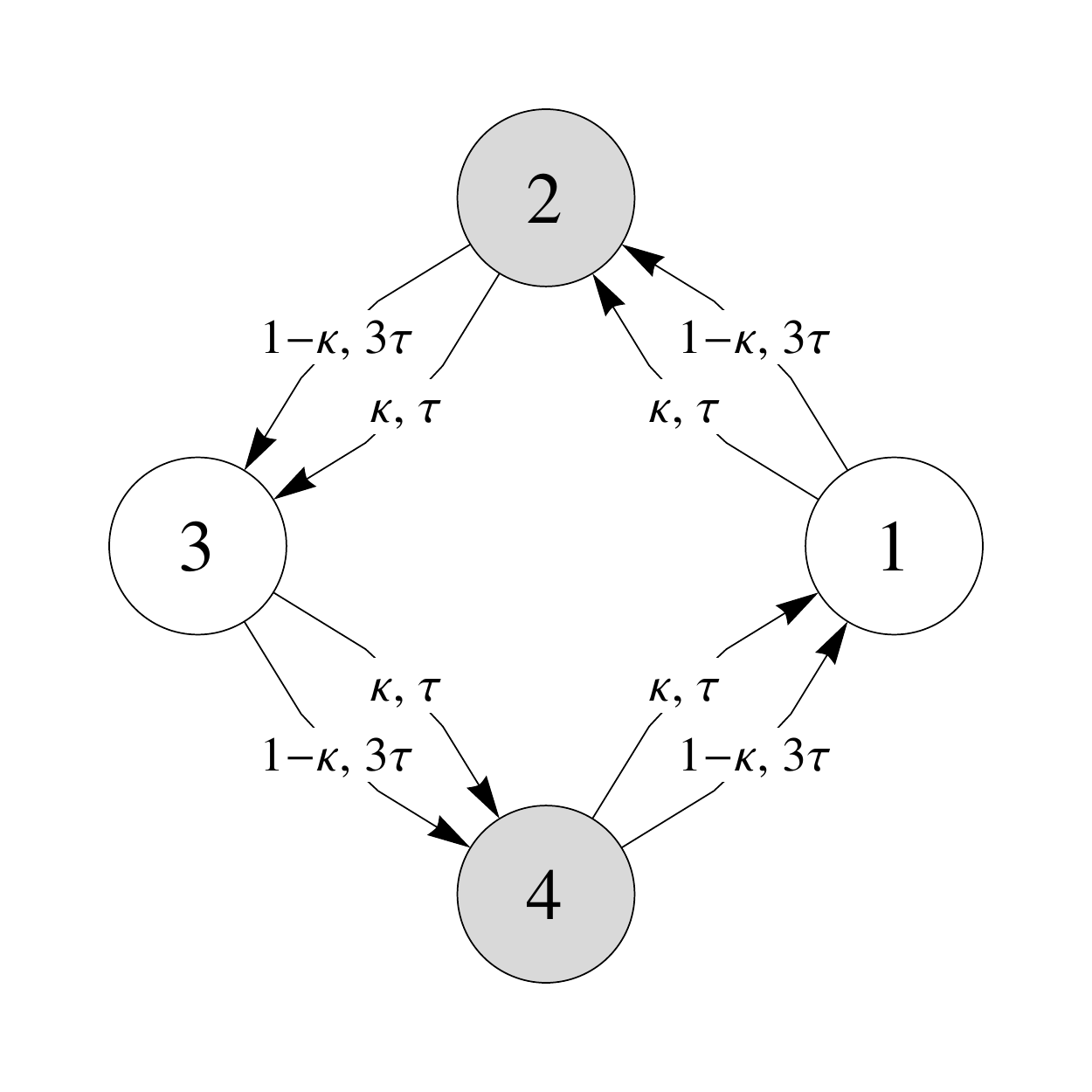}
    }
    \subfloat[\label{fig:4ringmap}]{
    \includegraphics[width=0.48\linewidth]{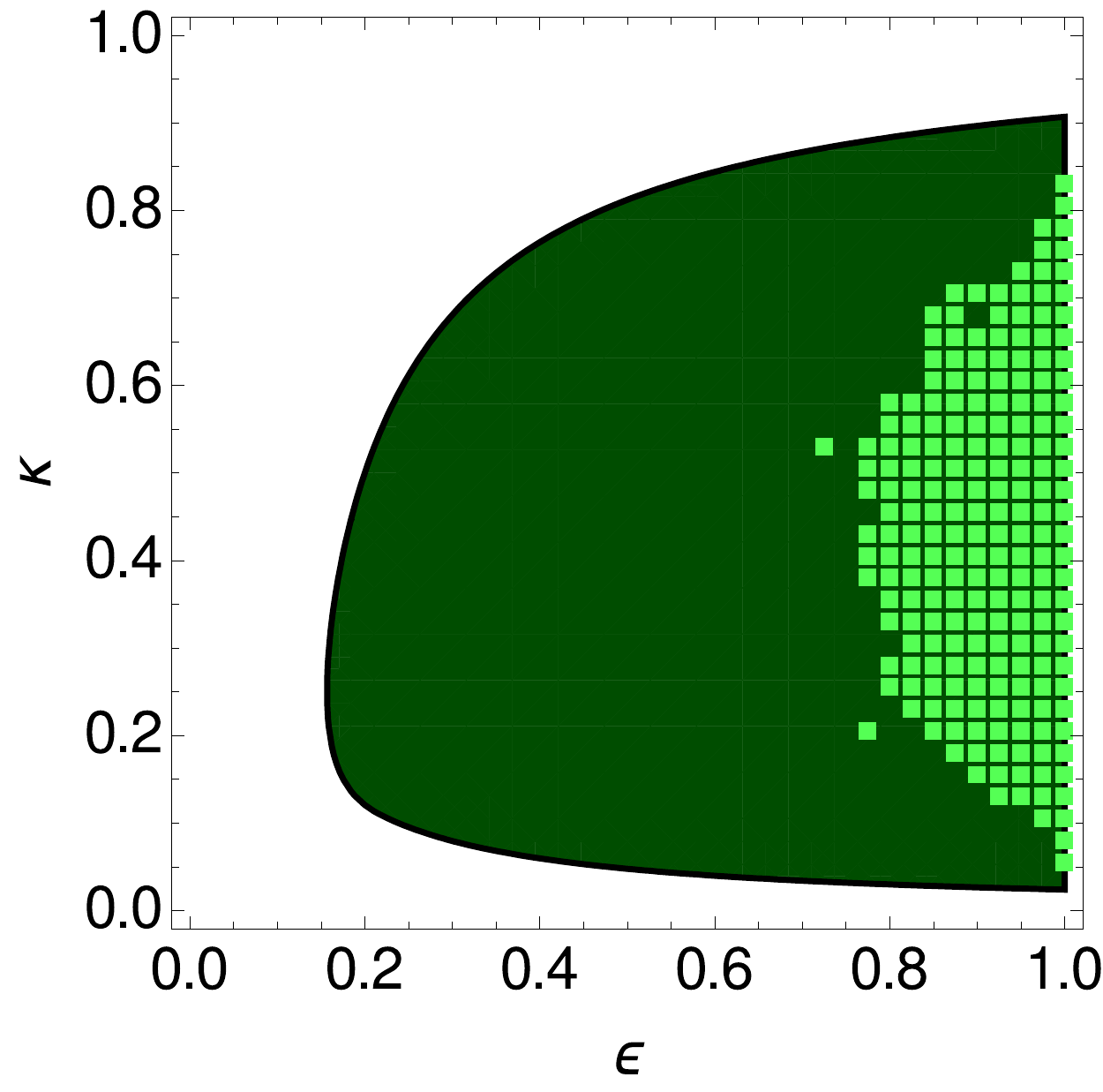}
    }
\caption{(color online). 
\protect\subref{fig:4ring} 
4 unit directed ring with 2 delay times $3\tau_1=\tau_2=100$. 
Units 1 and 3, and 2 and 4 respectively, belong to synchronized sublattices.
\protect\subref{fig:4ringmap}
Sublattice synchronization region for Bernouilli maps with $a=1.05$ obtained by solving Eq. \ref{eq:YPoly} 
numerically for modes 2 and 3, $\gamma_{2,3}=\pm i$, and intersecting their stability regions. Squares
mark stable sublattice synchronization for equivalent Tent maps, obtained by simulation
as in Fig \ref{fig:ringmapstents}.}
\label{fig:Sublattice4Ring}
\end{figure}

For a delay ratio of $p/q=1/3$ the resulting network contains loops of length $(N+2j)\tau$. The $GCD$ is equal 
to 2 for rings with an even number of units, and to 1 if the number of nodes is odd.
Hence, for even $N$, this time delay ratio produces sublattice synchronization.
As shown in Fig. \ref{fig:MSF}(b),
the master stability function is  symmetric under a transformation $\gamma\rightarrow -\gamma$. 
A transverse mode with eigenvalue $\gamma_n=-1$ is thus always unstable for chaotic dynamics. 
We provide an example by solving the specific case of a 4 unit doubly coupled ring (Fig. \ref{fig:4ring}).
Here, eigenmode $\vec{\omega_2}=[1,-1,1-1]$ with eigenvalue $\lambda_1=-1$ is unstable, generating two groups 
formed by units 1 and 3, and 2 and 4 respectively. The modes $\gamma_{1,3}=\pm i$, which do not 
allow sublattice synchronization, are both stable. Units belonging to the same group develop identical
chaotic trajectories, although they are not directly coupled, but receive input from units of the other group.
The resulting parameter region showing stable sublattice synchronization for Bernouill maps is shown in Fig. \ref{fig:4ringmap}.
In order to show that the phenomenon is not limited to constant slope maps, we 
also provide the synchronization region for an equivalent network of Tent maps. We observe
again a shrinking of the synchronization region due to time derivative fluctuations.

\subsection{Comparison with analogous single delay networks}\label{s:double_element}
\begin{figure}[h!]
    \subfloat[\label{fig:aux1}]{
    \includegraphics[width=0.48\linewidth]{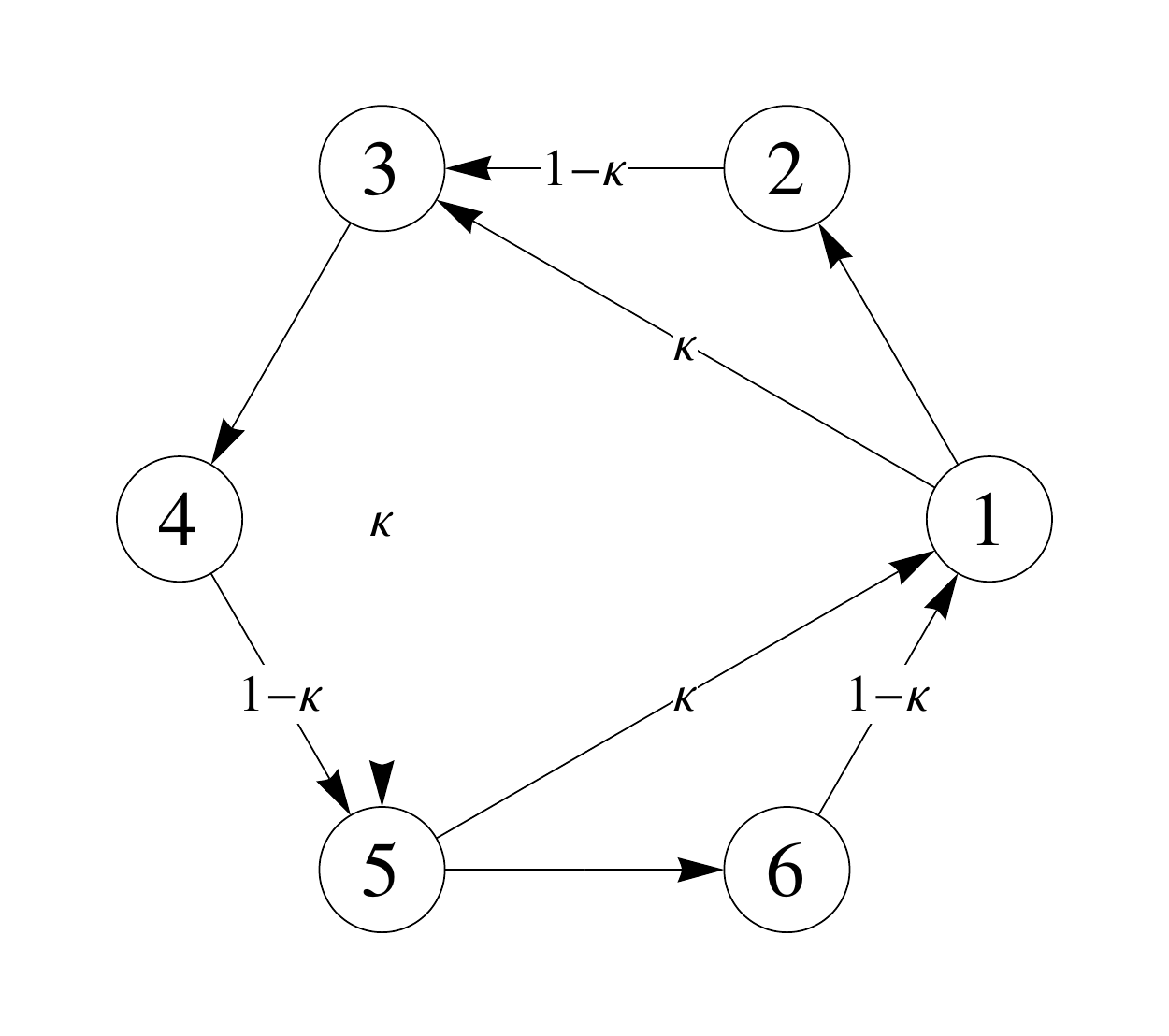}
    }
    \subfloat[\label{fig:aux2}]{
    \includegraphics[width=0.48\linewidth]{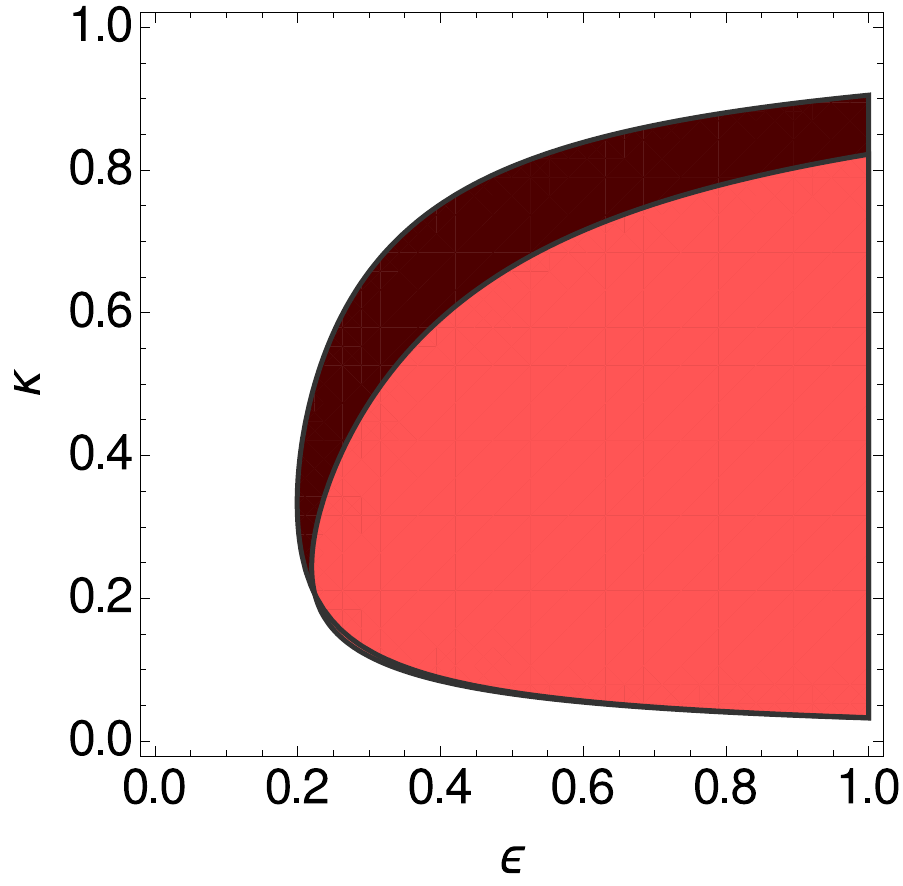}
    }
\caption{(color online). 
\protect\subref{fig:aux1} 
3 unit ring with auxiliary units analogous to the network depicted in Fig. \ref{fig:ringmaps}. 
The double delay $2\tau_1$ is
substituted by two links of delay length $\tau_1$ mediated by auxiliary units.
\protect\subref{fig:aux2}
The synchronization region, in light red, was obtained by solving Eq. \eqref{eq:Single} ($a=1.05$). 
It is smaller than the corresponding from the double delay case, in darker red, identical to the one 
shown in Fig. \ref{fig:ringmaps1}.}
\label{fig:auxiliaryunits}
\end{figure}

The effect of time delay resonances cannot be explained by the GCD condition alone.
In reference \cite{Nixon2012}, the effect of multiple time delay is studied
by transforming the network to an equivalent network with homogeneous delay times.
This is done by inserting imaginary units 
coupled with a single time delay along the longer connections. 
Here, we demonstrate how the resulting phase maps are different, despite both networks being equivalent 
from the GCD point of view.
Take for instance a directed ring of three units with $p/q=1/2$. 
With respect to the GCD condition, this network is completely analogous to a single delay triangle where the
longer links have been substituted by a 2 link chain of simple delays mediated by an auxiliary unit
(see Fig. \ref{fig:auxiliaryunits}). The corresponding synchronization region turns out to be smaller than that
of a directed triangle with double delay. Moreover, a single delay network like this one does not
suffer from detuning effects.

\subsection{Sensitivity to detuning}\label{s:detuning}

The synchronization properties of chaotic networks with multiple time-delay
depend on a precise ratio between the time delays. The synchronization phase maps presented above 
are found to be very sensitive to detuning. As it is shown
in Fig. \ref{fig:ringmaps2}, the parameter regions showing synchronization stability shrink drastically  
after a small mismatch of the delays, $\tau_1\rightarrow \tau_1-1$ and $\tau_2\rightarrow \tau_2-1$.

\begin{figure}[ht!]
\centering
    \subfloat[\label{fig:SL1}]{
    \includegraphics[width=0.5\linewidth]{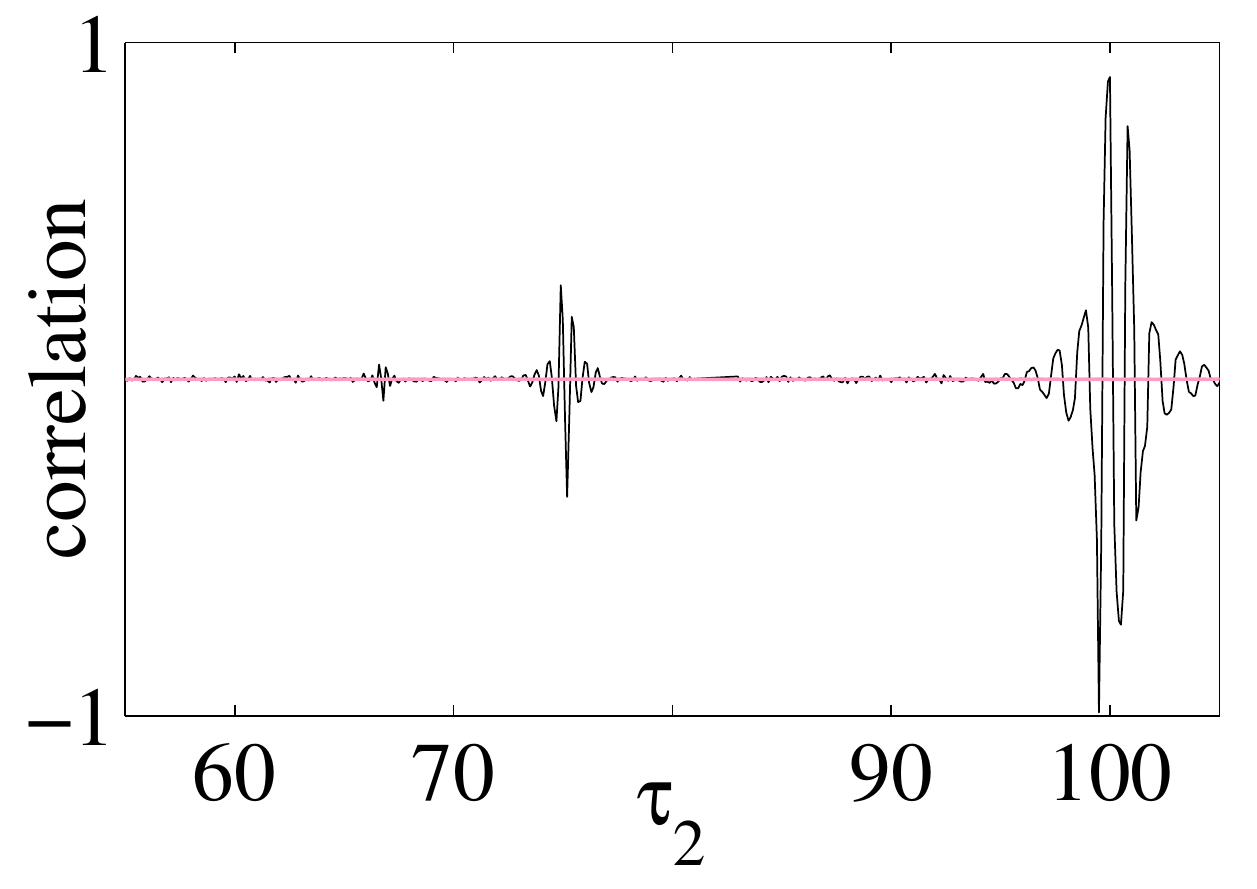} %
    }
    \subfloat[\label{fig:SL2}]{
    \includegraphics[width=0.5\linewidth]{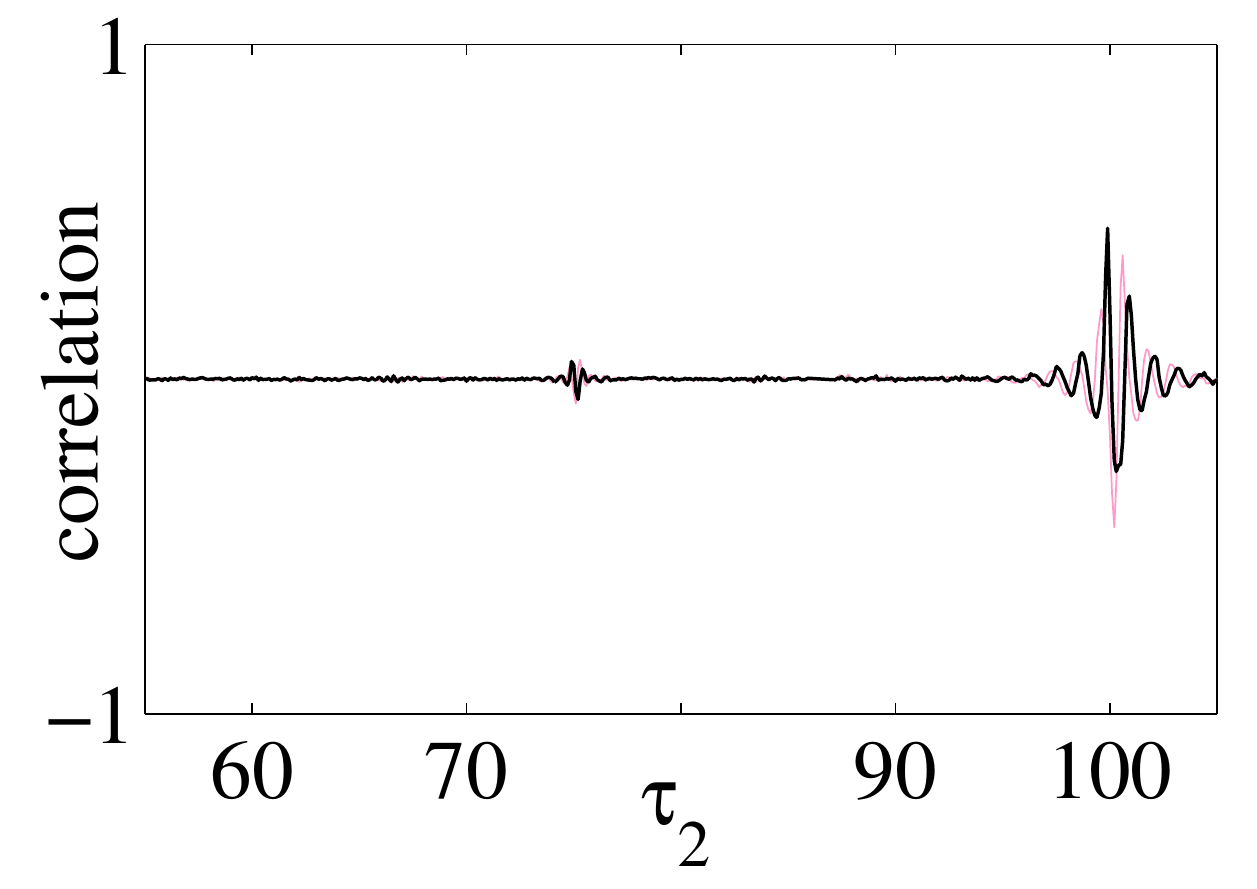}
    }
\caption{(color online). 
Correlation peaks corresponding to time delay resonances for (a) 2 mutually coupled
 and (b) a doubly coupled directed ring of 3 chaotic Stuart-Landau oscillators with $\tau_1=50$.
}
\label{fig:SL}
\end{figure}

It is more natural to study the effect of detuning between the two delay times in continuous chaotic systems 
with larger internal correlation time. Therefore, we consider a doubly connected unidirectional ring of Stuart-Landau oscillators, 
modeled by
\begin{multline}
\dot{z}_k(t)=z_k(t)(1-|z_k(t)|^2)+i\beta z_k(t)|z_k(t)|^2 \\
+\kappa_1 z_{k+1}(t-\tau_1)+\kappa_2 z_{k+1}(t-\tau_2)\label{eq:SL}\,,
\end{multline}
where the last unit, $k=N-1$, is coupled to the first, $k=0$. Here $\beta$ is the amplitude phase-coupling, 
$\kappa_1$ and $\kappa_2$ are the coupling strengths and $\tau_1$ and $\tau_2$ are the coupling delays.
Without coupling, the oscillators are in a stable periodic orbit $z(t)=e^{i\beta t}$. 
In Fig. \ref{fig:SL}(a) we show the unit-to-unit zero lag crosscorrelation, given by 
$\langle z_i(t) z_{i+1}^*(t) \rangle/\sqrt{\langle|z_i(t)|^2|z_{i+1}(t)|^2\rangle_t}$, as a function of 
the ratio between the delay times $\tau_2/\tau_1$ in a ring of two oscillators. 
The unit-to-unit zero lag crosscorrelation in a three-unit ring is shown in Fig. \ref{fig:SL}(b).
Both ring configurations show high correlations around delay ratios of $\tau_2/\tau_1= 4/3,\; 3/2,\; 5/3$ and $2$, 
as predicted by the GCD-argument. 
We observe, firstly, that the correlations are higher for two than for three oscillators.
For the ratio $\tau_2/\tau_1=2$, this corresponds to the smaller transverse eigenvalue found for two Bernoulli maps, represented in
Fig. \ref{fig:MSF}. Secondly, we find higher correlations for simpler ratios.
The width of the delay resonance peaks depends on the internal decay time of the oscillators: 
while the Bernouilli maps are found to be very sensitive to detuning of the two delay times, we find a considerable width of the 
resonances for Stuart-Landau oscillators. Moreover, the crosscorrelation between the oscillators is not always positive, but has an 
oscillatory shape as the delay ratio varies. This can be explained by phase effects: for two coupled oscillators 
we find anti-synchrony $z_1(t)\approx -z_2(t)$ for delay ratios $p\tau_2=q\tau_1 \pm \pi/\omega$, with $\omega$ being the dominant 
frequency of the chaotic motion. Indeed, having $z(t+\pi/\omega)\approx -z(t)$ 
one finds that for these delay ratios the synchronization manifold is destabilized, as the two delayed signals 
interfere destructively with each other. The manifold $z_1(t)=-z_2(t)$ is however not suppressed. Therefore 
we observe anti-correlation between the oscillators. Consequently, we find a frequency of $p\omega$ at the delay resonance 
$\tau_2/\tau_1=q/p$, i.e. the first order resonance at $\tau_2=2\tau_1$ has an oscillation frequency of 
$\omega\approx \beta$, the second order resonance at $2\tau_2=3\tau_1$ has a frequency of approximately $2\beta$, etc. 
Similar results can be found in the ring of three doubly connected oscillators.
\section{Non-zero correlations without synchronization}\label{s:corrs}
The GCD argument provides information about the possible  number of synchronization groups. 
This holds for networks where complete synchronization is a solution of the dynamical equations. 
But it was argued in reference \cite{Heiligenthal2011} that the GCD condition is applicable to other networks as 
well. Even if chaos synchronization is not a solution, the GCD determines 
how the information about the trajectories mixes according to the network topology. 
We demonstrate this in two different systems. First we show how the GCD affects correlations among the 
non-linear units in an asymmetrical network. 
Secondly, we study correlations in directed rings of coupled linear oscillators with noise, identifying correlation
peaks for the time delay ratios predicted by the GCD condition.
%
\subsection{Correlations in asymmetric network}\label{s:finitecorr}
\begin{figure}[t!]
    \subfloat[\label{fig:detour1}]{
      \includegraphics[width=0.4\linewidth]{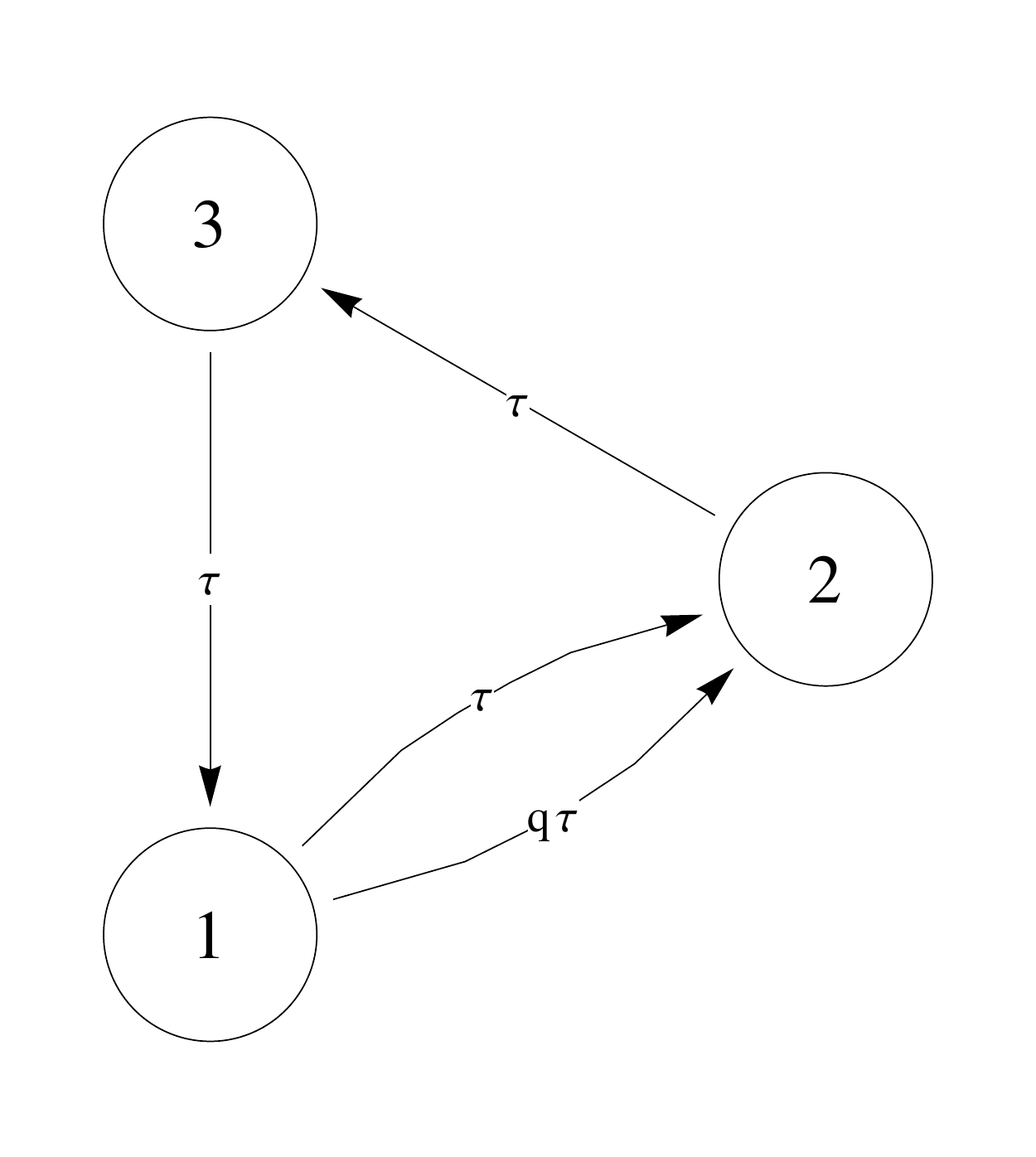}
    }
    \subfloat[\label{fig:detour2}]{
      \includegraphics[width=0.45\linewidth]{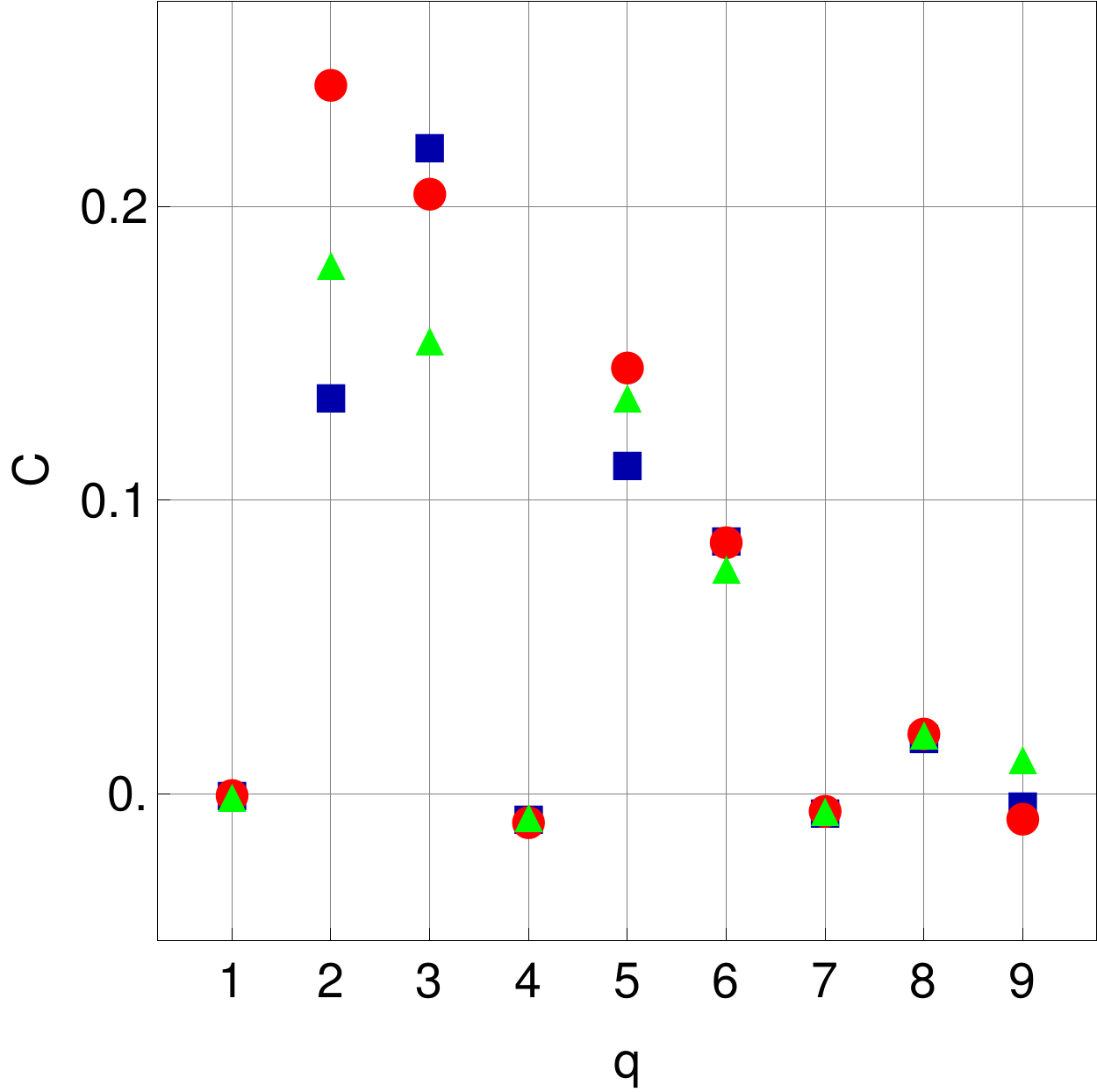}
    }
\caption{(color online). 
\protect\subref{fig:detour1} 
Non-synchronizable assymetric network of Bernouilli maps.
\protect\subref{fig:detour2}
Finite correlations among units: $C(1,2)$ 
(green triangles), $C(2,3)$ (blue squares) and $C(1,3)$ (red circles) for $\epsilon=\kappa=0.85$, $\tau=50$ 
after a transitory of 40000 time-steps, averaged over 100 trials and over a time-window of 10000 time-steps.
}
\label{fig:Detour}
\end{figure}

When the network GCD is equal to $K>1$, at each time-step each unit is driven just by the units belonging
to one of the synchronized sublattices. On the other hand, when the GCD is equal to 1, the graph is aperiodic 
and after enough iterations of the dynamics each unit is being driven by the initial state of every other unit.
This topological effect is able to induce correlations among the network trajectories even when
the system does not have a synchronized solution.
Consider the network depicted in Fig. \ref{fig:Detour}a. Here, we have a single coupling 
detour between two units 
with variable time delay $q\tau$ embedded in a 3 unit ring with single a time delay $\tau$. 
This network cannot synchronize by construction. We have simulated the dynamics of 
Bernouilli maps coupled with this topology and
computed time-correlations among each unit trajectories. The results are shown in Fig. \ref{fig:Detour}b.
For integer values of $q$, we can distinguish two situations.
For values of $q=2,3,5,6,8,9$, we have $GCD=1$.  
However, for $q=1,4,7$ we have $GCD=3$ and the correlations are practically zero. 

\subsection{Linear systems}\label{s:linearsys}
It is useful to compare chaotic systems to linear systems with noise. By replacing the chaotic 
dynamics by a linear system with white noise, one recovers properties that relate solely 
to network structure, and not to the specific chaotic system. However, a major difference between 
deterministic chaotic systems and stochastic linear systems is that the latter cannot synchronize, 
as synchronization is a nonlinear phenomenon. Nevertheless, linear stochastic systems have been shown 
to mimic several qualitative features of the auto- and cross-correlation functions of delay-coupled 
chaotic elements. In some cases one can even quantitatively model the auto-correlation function of a 
chaotic delay-system with a linear model \cite{DHuys2012}. We demonstrate here that also these delay 
resonances can be explained by a stochastic linear delay model.

We consider a ring of $N$ oscillators, where each oscillator is characterized by a natural 
frequency $\omega_0$, an internal decay rate $\alpha$, and an internal white gaussian noise 
source $\xi_k(t)$, with zero mean (the variance is irrelevant, as the whole system can be rescaled).
We thus approximate the chaotic signal by a linear response, which is captured by the correlation
functions, and a component which effectively acts as a source of noise. Each node is coupled to its 
neighbor with a strength $\kappa_1$ over a first connection with a delay $\tau_1$, and a second 
connection with a delay $\tau_2$ and a strength $\kappa_2$. This system is modeled as
\begin{eqnarray}
\dot{x}_k &=& (-\alpha+i\omega_0) x_k+\xi_k(t)\nonumber\\
& & +\kappa_1 x_{k+1}(t-\tau_1)+\kappa_2 x_{k+1}(t-\tau_2)\,,
\label{eq:lin}
\end{eqnarray}
with $k=N\equiv0$. We can decompose the system into its eigenmodes 
$v_n(t)$, given by $$v_n=\frac{1}{\sqrt{N}}\displaystyle\sum_{k=0}^{N-1}e^{ink\theta}x_k,$$
with $\theta=2\pi/N$. The dynamics is then modeled by 
\begin{eqnarray}
\dot{v}_n &=& (-\alpha+i\omega_0) v_n+\xi_n(t)\nonumber \\
& & +\kappa_1 e^{in\theta} v_n(t-\tau_1)+\kappa_2 e^{in\theta} v_n(t-\tau_2)\,,
\end{eqnarray}
with $\xi_n(t)=1/\sqrt{N}\sum e^{ink\theta}\xi_k(t)$. We can easily solve the system in Fourier space, 
and find for $\mathcal{F}(v_n(t))=\tilde{v}_n(\omega)$
\begin{equation}
\tilde{v}_n(\omega)=\frac{\tilde{\xi}_n(\omega)}{\alpha+i(\omega-\omega_0)-\kappa_1 e^{in\theta-i\omega\tau_1}-\kappa_2 e^{in\theta-i\omega\tau_2}}\,.
\label{eq:spec}
\end{equation}
The spectrum $\tilde{v}_n(\omega)$ has maxima at $\omega\tau_1=2m\pi+n\theta$ and at 
$\omega\tau_2=2l\pi+n\theta$; the peaks are most pronounced when those two conditions are both hold. 

\begin{figure}[b!]
    \subfloat[\label{fig:autocov}]{
    \includegraphics[width=0.5\linewidth]{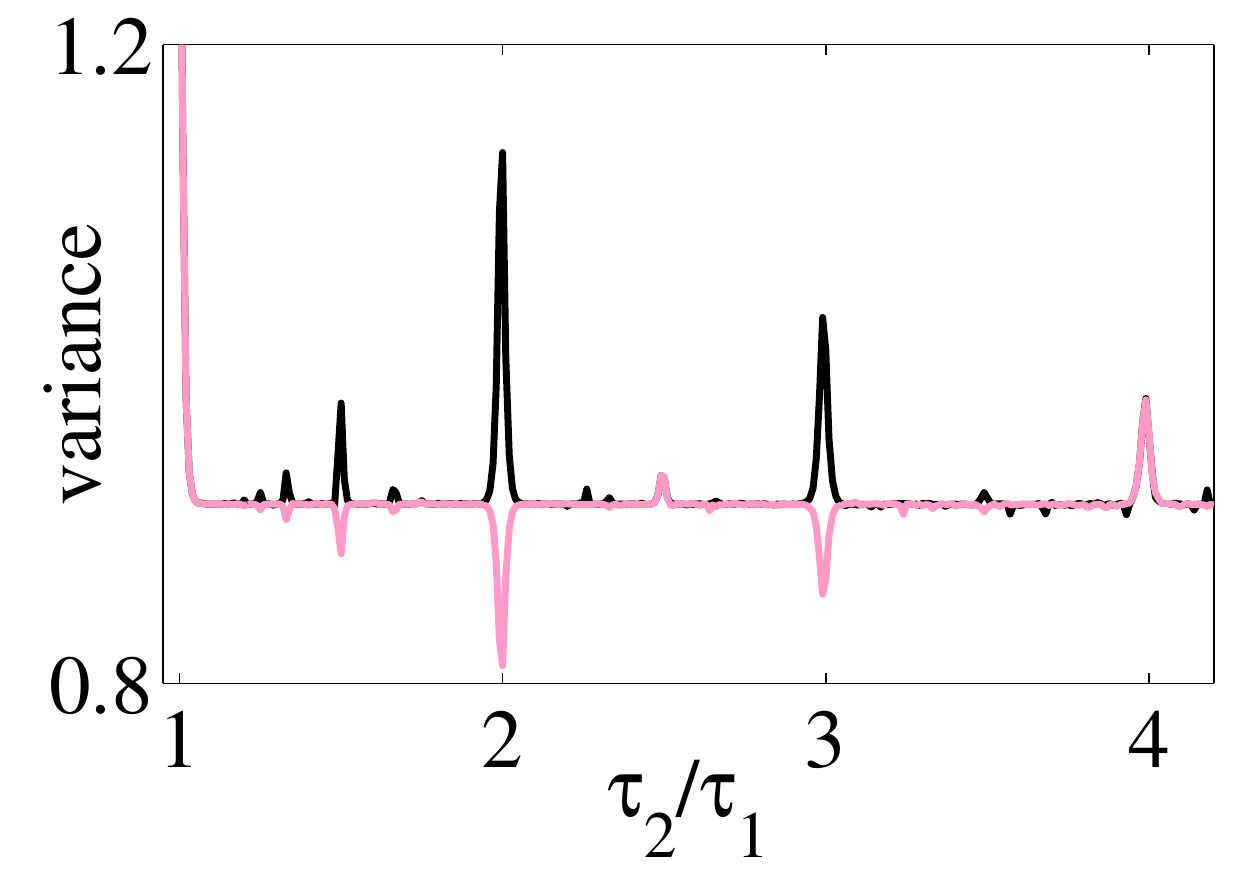} %
    }
    \subfloat[\label{fig:crossco}]{
    \includegraphics[width=0.5\linewidth]{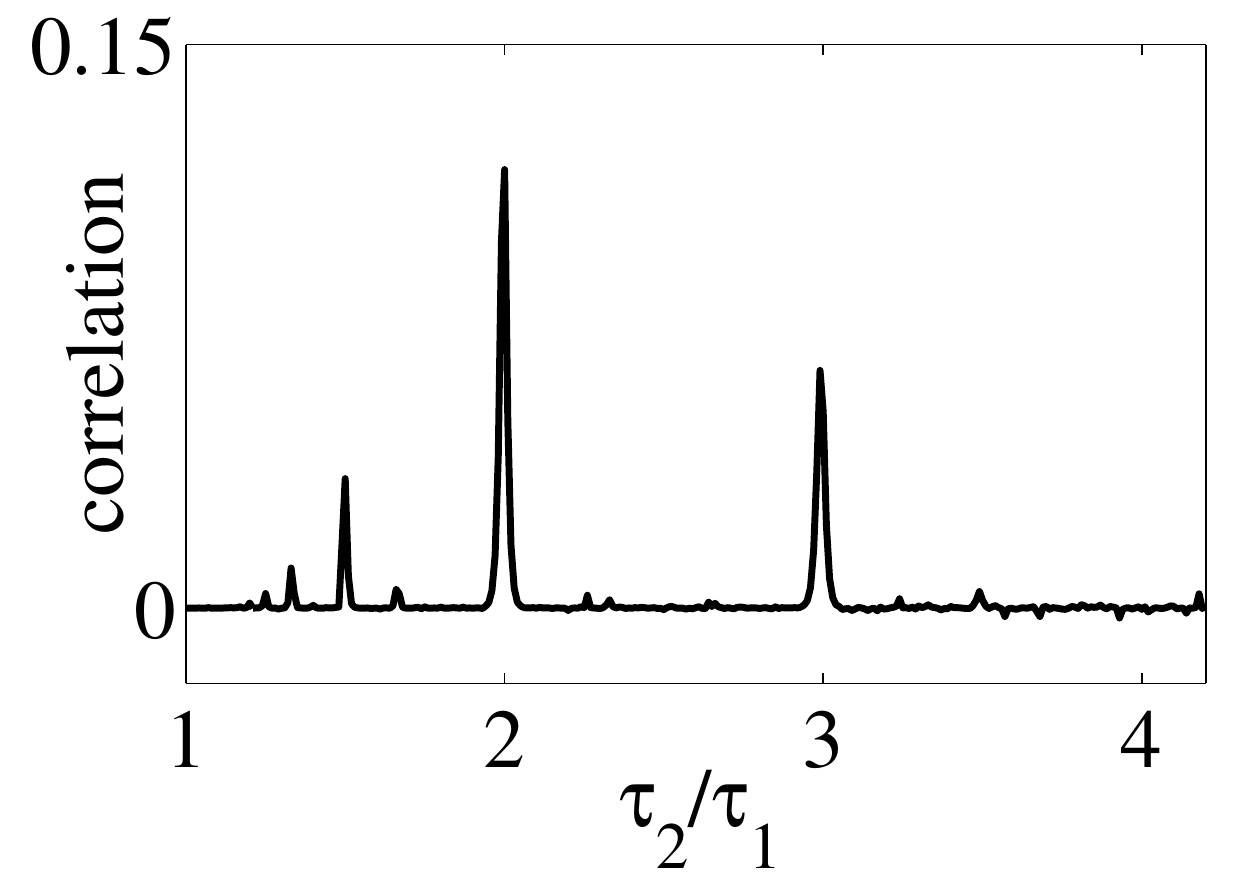}
    }
\caption{(color online). 
\protect\subref{fig:autocov} Variances of the in-phase eigenmode $\langle|v_0(t)|^2\rangle$ (black curve)
and the first out-of-phase eigenmode $\langle|v_1(t)|^2\rangle$ (pink curve)
in a ring of 3 linear noisy oscillators coupled with two delays 
(Eq. \eqref{eq:lin}) as function of the ratio of the delays.
\protect\subref{fig:crossco} Node-to-node correlation in a ring of 
three nodes with two delays, as a function of the ratio of the delays. 
Parameters are $\kappa_1=\kappa_2=2.25$, $\tau_1=50$, $\omega_0=0$ and 
$\langle \xi^2(t)\rangle=1$.
}
\label{fig:LinearSystems}
\end{figure}

Clearly, the variances 
$$\langle |v_n(t)|^2\rangle=\displaystyle\int_{-\infty}^{+\infty}d\omega |\tilde{v}_n(\omega)|$$ 
are maximal when the resonances due to the two delays maximally overlap, and minimal when none of the 
resonances overlap. Consequently, the in-phase mode $\tilde{v}_0$ has a maximal variance for $\tau_2/\tau_1$ being 
rational; the simpler this ratio, the larger the variance. The out-of-phase modes $n>0$ have a larger 
variance for $\tau_2/\tau_1=(n+lN)/(n+mN)$ and a minimal variance for rational ratios for which this 
conditions does not hold. 

We show the variances of the different modes in a ring of $N=3$ elements in Fig. \ref{fig:LinearSystems}a. 
The in-phase mode has large maxima at $\tau_2=l\tau_1$, and less pronounced maxima at other rational ratios as 
$\tau_2/\tau_1=6/5,4/3,3/2,5/3,7/4, 5/2,\hdots$. The out-of-phase mode shows maxima 
if $\tau_2/\tau_1=(3l+1)/(3m+1)$, we see indeed extrema at $\tau_2/\tau_1=1, 5/2, 4$ and minima 
for $\tau_2=2\tau_1$ and $\tau_2=3\tau_1$.
The crosscorrelation between two nodes in the ring, $\langle x_0(t) x_k^*(t) \rangle_t$ can be computed 
as a sum of the eigenmodes:
\begin{eqnarray}
\langle x_0(t) x_k^*(t) \rangle_t & = & \mathcal{F}^{-1}\left(\tilde{x}_0(\omega)\tilde{x}_k^*(\omega)\right)\nonumber\\
& = &\frac{1}{N} \displaystyle\int^{+\infty}_{-\infty} d\omega \sum_n \tilde{v}_n(\omega) \sum_m e^{imk\theta}\tilde{v}_m^*(\omega)\nonumber\\
& = &\frac{1}{N} \sum_n e^{ink\theta} \int^{+\infty}_{-\infty} d\omega |\tilde{v}_n(\omega)|^2
\end{eqnarray}
If all the eigenmodes $v_n(t)$ have the same variance the sum cancels out and there will not be any 
zero lag correlation  between the two elements. This is the case if $\tau_2/\tau_1$ is irrational 
and if multiple eigenmodes have overlapping resonances for the same delay ratio. This happens whenever
\begin{equation}
\frac{\tau_2}{\tau_1}= \frac{n+lN}{n+mN}\,,
\end{equation}
for a given mode $1\leq n<N$, effectively recovering Eq. \ref{eq:forbidden_ratio}. \
For sublattice synchronization, it should hold for all $k$, except for the distance 
$K$ between the nodes of the group.
For a ring of $N=3$ elements, we show the magnitude of the zero-lag correlation in 
Fig. \ref{fig:LinearSystems}b.
%
%
\section{Summary}\label{s:summ}
We have provided a formalism for studying the stability of synchronization in networks with two or more time delays.
Some networks that do not synchronize with a single time delay can be brought to synchrony with an appropiate choice of the 
second time delay. Ultimately, the relationship between both time delays determines the available 
synchronization regime, as predicted by the GCD condition. We exemplified this comprehensively in the case of directed 
rings with two delay times. We were able to provide certain time delay ratios that do not allow complete synchronization,
as well as to identify the unstable modes giving rise to synchronized sublattices.
The resulting synchronization stability regions are different from those in GCD equivalent networks with single time delay.
Also, the synchronization properties can be very sensitive to a detuning between the two time delays: as one of the time delays varies, 
we observe positive or negative correlations depending on the internal correlation time and frequency of the chaotic units.

The global influence of network topology and time delay resonance via the GCD argument is made evident not only
in synchronization phenomena. The network structure also induces high correlations among the units trajectories in 
non-synchronizable systems. Moreover, the GCD-induced time delay resonances observed in rings of chaotic map are 
reproduced by networks of noisy linear oscillators. Yet, a mathematically rigorous explanation of the GCD argument 
in general chaotic networks remains to be found. 
\begin{acknowledgments} 
  M. Jim\'enez and E. Korutcheva warmly thank the hospitality and the financial support of 
  Dep. of Theoretical Physics III at the University of Würzburg, as well as the financial
  support of Dep. de F\'isica Fundamental, UNED. E. Korutcheva also thanks the sponsorship 
  by the Alexander von Humboldt Foundation within the Renewed research stay program.
\end{acknowledgments}
%

\begin{thebibliography}{29}
\expandafter\ifx\csname natexlab\endcsname\relax\def\natexlab#1{#1}\fi
\expandafter\ifx\csname bibnamefont\endcsname\relax
  \def\bibnamefont#1{#1}\fi
\expandafter\ifx\csname bibfnamefont\endcsname\relax
  \def\bibfnamefont#1{#1}\fi
\expandafter\ifx\csname citenamefont\endcsname\relax
  \def\citenamefont#1{#1}\fi
\expandafter\ifx\csname url\endcsname\relax
  \def\url#1{\texttt{#1}}\fi
\expandafter\ifx\csname urlprefix\endcsname\relax\def\urlprefix{URL }\fi
\providecommand{\bibinfo}[2]{#2}
\providecommand{\eprint}[2][]{\url{#2}}

\bibitem[{\citenamefont{Boccaletti et~al.}(2002)\citenamefont{Boccaletti,
  Kurths, Osipov, Valladares, and Zhou}}]{Boccaletti2002}
\bibinfo{author}{\bibfnamefont{S.}~\bibnamefont{Boccaletti}},
  \bibinfo{author}{\bibfnamefont{J.}~\bibnamefont{Kurths}},
  \bibinfo{author}{\bibfnamefont{G.}~\bibnamefont{Osipov}},
  \bibinfo{author}{\bibfnamefont{D.}~\bibnamefont{Valladares}},
  \bibnamefont{and} \bibinfo{author}{\bibfnamefont{C.}~\bibnamefont{Zhou}},
  \bibinfo{journal}{Physics Reports} \textbf{\bibinfo{volume}{366}},
  \bibinfo{pages}{1 } (\bibinfo{year}{2002}), ISSN \bibinfo{issn}{0370-1573}.

\bibitem[{\citenamefont{Pikovsky et~al.}(2001)\citenamefont{Pikovsky,
  Rosenblum, and Kurths}}]{Pikovsky2001}
\bibinfo{author}{\bibfnamefont{A.}~\bibnamefont{Pikovsky}},
  \bibinfo{author}{\bibfnamefont{M.~G.} \bibnamefont{Rosenblum}},
  \bibnamefont{and} \bibinfo{author}{\bibfnamefont{J.}~\bibnamefont{Kurths}},
  \emph{\bibinfo{title}{Synchronization, A Universal Concept in Nonlinear
  Sciences}} (\bibinfo{publisher}{Cambridge University Press},
  \bibinfo{address}{Cambridge}, \bibinfo{year}{2001}).

\bibitem[{\citenamefont{Erneux}(2009)}]{Erneux2009}
\bibinfo{author}{\bibfnamefont{T.}~\bibnamefont{Erneux}},
  \emph{\bibinfo{title}{Applied Delay Differential Equations}},
  vol.~\bibinfo{volume}{3} of \emph{\bibinfo{series}{Surveys and Tutorials in
  the Applied Mathematical Sciences}} (\bibinfo{publisher}{Springer-Verlag New
  York}, \bibinfo{year}{2009}), ISBN \bibinfo{isbn}{978-0-387-74371-4}.

\bibitem[{\citenamefont{Atay et~al.}(2004)\citenamefont{Atay, Jost, and
  Wende}}]{Atay2004}
\bibinfo{author}{\bibfnamefont{F.~M.} \bibnamefont{Atay}},
  \bibinfo{author}{\bibfnamefont{J.}~\bibnamefont{Jost}}, \bibnamefont{and}
  \bibinfo{author}{\bibfnamefont{A.}~\bibnamefont{Wende}},
  \bibinfo{journal}{Phys. Rev. Lett.} \textbf{\bibinfo{volume}{92}},
  \bibinfo{pages}{144101} (\bibinfo{year}{2004}).

\bibitem[{\citenamefont{Lakshmanan and Senthilkumar}(2011)}]{Lakshmanan2011}
\bibinfo{author}{\bibfnamefont{M.}~\bibnamefont{Lakshmanan}} \bibnamefont{and}
  \bibinfo{author}{\bibfnamefont{D.~V.} \bibnamefont{Senthilkumar}},
  \emph{\bibinfo{title}{Dynamics of Nonlinear Time-Delay Systems}}
  (\bibinfo{publisher}{Springer Berlin Heidelberg}, \bibinfo{year}{2011}), ISBN
  \bibinfo{isbn}{978-3-642-14937-5 (Print) 978-3-642-14938-2 (Online)}.

\bibitem[{\citenamefont{Murphy et~al.}(2009)\citenamefont{Murphy, Cohen,
  Ravoori, Schmitt, Setty, Sorrentino, Williams, Ott, and Roy}}]{Murphy2009}
\bibinfo{author}{\bibfnamefont{T.~E.} \bibnamefont{Murphy}},
  \bibinfo{author}{\bibfnamefont{A.~B.} \bibnamefont{Cohen}},
  \bibinfo{author}{\bibfnamefont{B.}~\bibnamefont{Ravoori}},
  \bibinfo{author}{\bibfnamefont{K.~R.~B.} \bibnamefont{Schmitt}},
  \bibinfo{author}{\bibfnamefont{A.~V.} \bibnamefont{Setty}},
  \bibinfo{author}{\bibfnamefont{F.}~\bibnamefont{Sorrentino}},
  \bibinfo{author}{\bibfnamefont{C.~R.~S.} \bibnamefont{Williams}},
  \bibinfo{author}{\bibfnamefont{E.}~\bibnamefont{Ott}}, \bibnamefont{and}
  \bibinfo{author}{\bibfnamefont{R.}~\bibnamefont{Roy}},
  \bibinfo{journal}{Philosophical Transactions of the Royal Society of London
  A: Mathematical, Physical and Engineering Sciences}
  \textbf{\bibinfo{volume}{368}}, \bibinfo{pages}{343} (\bibinfo{year}{2009}),
  ISSN \bibinfo{issn}{1364-503X}.

\bibitem[{\citenamefont{Locquet et~al.}(2002)\citenamefont{Locquet, Masoller,
  and Mirasso}}]{Mirasso2002}
\bibinfo{author}{\bibfnamefont{A.}~\bibnamefont{Locquet}},
  \bibinfo{author}{\bibfnamefont{C.}~\bibnamefont{Masoller}}, \bibnamefont{and}
  \bibinfo{author}{\bibfnamefont{C.~R.} \bibnamefont{Mirasso}},
  \bibinfo{journal}{Phys. Rev. E} \textbf{\bibinfo{volume}{65}},
  \bibinfo{pages}{056205} (\bibinfo{year}{2002}).

\bibitem[{\citenamefont{Fischer et~al.}(2006)\citenamefont{Fischer, Vicente,
  Buldu, Peil, Mirasso, Torrent, and Garcia-Ojalvo}}]{Fischer06}
\bibinfo{author}{\bibfnamefont{I.}~\bibnamefont{Fischer}},
  \bibinfo{author}{\bibfnamefont{R.}~\bibnamefont{Vicente}},
  \bibinfo{author}{\bibfnamefont{J.M.}~\bibnamefont{Buldu}},
  \bibinfo{author}{\bibfnamefont{M.}~\bibnamefont{Peil}},
  \bibinfo{author}{\bibfnamefont{C.R.}~\bibnamefont{Mirasso}},
  \bibinfo{author}{\bibfnamefont{M.C.}~\bibnamefont{Torrent}}, \bibnamefont{and}
  \bibinfo{author}{\bibfnamefont{J.}~\bibnamefont{Garcia-Ojalvo}},
  \bibinfo{journal}{Phys. Rev. Lett.} \textbf{\bibinfo{volume}{97}},
  \bibinfo{pages}{123902} (\bibinfo{year}{2006}).

\bibitem[{\citenamefont{Peil et~al.}(2007)\citenamefont{Peil, Larger, and
  Fischer}}]{PeilFischer07}
\bibinfo{author}{\bibfnamefont{M.}~\bibnamefont{Peil}},
  \bibinfo{author}{\bibfnamefont{L.}~\bibnamefont{Larger}}, \bibnamefont{and}
  \bibinfo{author}{\bibfnamefont{I.}~\bibnamefont{Fischer}},
  \bibinfo{journal}{Phys. Rev. E} \textbf{\bibinfo{volume}{76}},
  \bibinfo{eid}{045201} (pages~\bibinfo{numpages}{4}) (\bibinfo{year}{2007}).

\bibitem[{\citenamefont{Soriano et~al.}(2013)\citenamefont{Soriano,
  Garc\'{\i}a-Ojalvo, Mirasso, and Fischer}}]{Soriano2013}
\bibinfo{author}{\bibfnamefont{M.~C.} \bibnamefont{Soriano}},
  \bibinfo{author}{\bibfnamefont{J.}~\bibnamefont{Garc\'{\i}a-Ojalvo}},
  \bibinfo{author}{\bibfnamefont{C.~R.} \bibnamefont{Mirasso}},
  \bibnamefont{and} \bibinfo{author}{\bibfnamefont{I.}~\bibnamefont{Fischer}},
  \bibinfo{journal}{Rev. Mod. Phys.} \textbf{\bibinfo{volume}{85}},
  \bibinfo{pages}{421} (\bibinfo{year}{2013}).

\bibitem[{\citenamefont{Buzsaki}(2006)}]{Buzsaki2009}
\bibinfo{author}{\bibfnamefont{G.}~\bibnamefont{Buzsaki}},
  \emph{\bibinfo{title}{Rhythms of the brain}} (\bibinfo{publisher}{Oxford
  University Press}, \bibinfo{year}{2006}), ISBN \bibinfo{isbn}{9780199828234}.

\bibitem[{\citenamefont{Keane et~al.}(2012)\citenamefont{Keane, Dahms, Lehnert,
  Suryanarayana, Hövel, and Schöll}}]{Keane2012}
\bibinfo{author}{\bibfnamefont{A.}~\bibnamefont{Keane}},
  \bibinfo{author}{\bibfnamefont{T.}~\bibnamefont{Dahms}},
  \bibinfo{author}{\bibfnamefont{J.}~\bibnamefont{Lehnert}},
  \bibinfo{author}{\bibfnamefont{S.}~\bibnamefont{Suryanarayana}},
  \bibinfo{author}{\bibfnamefont{P.}~\bibnamefont{Hövel}}, \bibnamefont{and}
  \bibinfo{author}{\bibfnamefont{E.}~\bibnamefont{Schöll}},
  \bibinfo{journal}{The European Physical Journal B}
  \textbf{\bibinfo{volume}{85}}, \bibinfo{eid}{407} (\bibinfo{year}{2012}),
  ISSN \bibinfo{issn}{1434-6028}.

\bibitem[{\citenamefont{Argyris et~al.}(2005)\citenamefont{Argyris, Syvridis,
  Larger, Annovazzi-Lodi, Colet, Fischer, García-Ojalvo, Mirasso, Pesquera,
  and Shore}}]{Shore2005}
\bibinfo{author}{\bibfnamefont{A.}~\bibnamefont{Argyris}},
  \bibinfo{author}{\bibfnamefont{D.}~\bibnamefont{Syvridis}},
  \bibinfo{author}{\bibfnamefont{L.}~\bibnamefont{Larger}},
  \bibinfo{author}{\bibfnamefont{V.}~\bibnamefont{Annovazzi-Lodi}},
  \bibinfo{author}{\bibfnamefont{P.}~\bibnamefont{Colet}},
  \bibinfo{author}{\bibfnamefont{I.}~\bibnamefont{Fischer}},
  \bibinfo{author}{\bibfnamefont{J.}~\bibnamefont{García-Ojalvo}},
  \bibinfo{author}{\bibfnamefont{C.~R.} \bibnamefont{Mirasso}},
  \bibinfo{author}{\bibfnamefont{L.}~\bibnamefont{Pesquera}}, \bibnamefont{and}
  \bibinfo{author}{\bibfnamefont{K.~A.} \bibnamefont{Shore}},
  \bibinfo{journal}{Nature} \textbf{\bibinfo{volume}{438}},
  \bibinfo{pages}{343} (\bibinfo{year}{2005}).

\bibitem[{\citenamefont{Heiligenthal et~al.}(2011)\citenamefont{Heiligenthal,
  Dahms, Yanchuk, J\"ungling, Flunkert, Kanter, Sch\"oll, and
  Kinzel}}]{Heiligenthal2011}
\bibinfo{author}{\bibfnamefont{S.}~\bibnamefont{Heiligenthal}},
  \bibinfo{author}{\bibfnamefont{T.}~\bibnamefont{Dahms}},
  \bibinfo{author}{\bibfnamefont{S.}~\bibnamefont{Yanchuk}},
  \bibinfo{author}{\bibfnamefont{T.}~\bibnamefont{J\"ungling}},
  \bibinfo{author}{\bibfnamefont{V.}~\bibnamefont{Flunkert}},
  \bibinfo{author}{\bibfnamefont{I.}~\bibnamefont{Kanter}},
  \bibinfo{author}{\bibfnamefont{E.}~\bibnamefont{Sch\"oll}}, \bibnamefont{and}
  \bibinfo{author}{\bibfnamefont{W.}~\bibnamefont{Kinzel}},
  \bibinfo{journal}{Phys. Rev. Lett.} \textbf{\bibinfo{volume}{107}},
  \bibinfo{pages}{234102} (\bibinfo{year}{2011}).

\bibitem[{\citenamefont{Flunkert et~al.}(2010)\citenamefont{Flunkert, Yanchuk,
  Dahms, and Sch\"oll}}]{Scholl2010}
\bibinfo{author}{\bibfnamefont{V.}~\bibnamefont{Flunkert}},
  \bibinfo{author}{\bibfnamefont{S.}~\bibnamefont{Yanchuk}},
  \bibinfo{author}{\bibfnamefont{T.}~\bibnamefont{Dahms}}, \bibnamefont{and}
  \bibinfo{author}{\bibfnamefont{E.}~\bibnamefont{Sch\"oll}},
  \bibinfo{journal}{Phys. Rev. Lett.} \textbf{\bibinfo{volume}{105}},
  \bibinfo{pages}{254101} (\bibinfo{year}{2010}).

\bibitem[{\citenamefont{Heil et~al.}(2001)\citenamefont{Heil, Fischer,
  Els{\"{a}}{\ss}er, Mulet, and Mirasso}}]{Heiligenthal2001}
\bibinfo{author}{\bibfnamefont{T.}~\bibnamefont{Heil}},
  \bibinfo{author}{\bibfnamefont{I.}~\bibnamefont{Fischer}},
  \bibinfo{author}{\bibfnamefont{W.}~\bibnamefont{Els{\"{a}}{\ss}er}},
  \bibinfo{author}{\bibfnamefont{J.}~\bibnamefont{Mulet}}, \bibnamefont{and}
  \bibinfo{author}{\bibfnamefont{C.R.}~\bibnamefont{Mirasso}},
  \bibinfo{journal}{Phys. Rev. Lett.} \textbf{\bibinfo{volume}{86}},
  \bibinfo{pages}{795} (\bibinfo{year}{2001}).

\bibitem[{\citenamefont{Buldu et~al.}(2007)\citenamefont{Buldu, Torrent, and
  Garcia-Ojalvo}}]{buldu2007}
\bibinfo{author}{\bibfnamefont{J.}~\bibnamefont{Buldu}},
  \bibinfo{author}{\bibfnamefont{M.}~\bibnamefont{Torrent}}, \bibnamefont{and}
  \bibinfo{author}{\bibfnamefont{J.}~\bibnamefont{Garcia-Ojalvo}},
  \bibinfo{journal}{Journal of Lightwave Technology}
  \textbf{\bibinfo{volume}{25}}, \bibinfo{pages}{1549} (\bibinfo{year}{2007}).

\bibitem[{\citenamefont{Kanter et~al.}(2011{\natexlab{a}})\citenamefont{Kanter,
  Zigzag, Englert, Geissler, and Kinzel}}]{KanterKinzel2011}
\bibinfo{author}{\bibfnamefont{I.}~\bibnamefont{Kanter}},
  \bibinfo{author}{\bibfnamefont{M.}~\bibnamefont{Zigzag}},
  \bibinfo{author}{\bibfnamefont{A.}~\bibnamefont{Englert}},
  \bibinfo{author}{\bibfnamefont{F.}~\bibnamefont{Geissler}}, \bibnamefont{and}
  \bibinfo{author}{\bibfnamefont{W.}~\bibnamefont{Kinzel}},
  \bibinfo{journal}{EPL (Europhysics Letters)} \textbf{\bibinfo{volume}{93}},
  \bibinfo{pages}{60003} (\bibinfo{year}{2011}{\natexlab{a}}).

\bibitem[{\citenamefont{Kestler et~al.}(2007)\citenamefont{Kestler, Kinzel, and
  Kanter}}]{KestlerKanter2007}
\bibinfo{author}{\bibfnamefont{J.}~\bibnamefont{Kestler}},
  \bibinfo{author}{\bibfnamefont{W.}~\bibnamefont{Kinzel}}, \bibnamefont{and}
  \bibinfo{author}{\bibfnamefont{I.}~\bibnamefont{Kanter}},
  \bibinfo{journal}{Phys. Rev. E} \textbf{\bibinfo{volume}{76}},
  \bibinfo{pages}{035202} (\bibinfo{year}{2007}).

\bibitem[{\citenamefont{Kanter et~al.}(2011{\natexlab{b}})\citenamefont{Kanter,
  Kopelowitz, Vardi, Zigzag, Kinzel, Abeles, and Cohen}}]{KanterCohen2011}
\bibinfo{author}{\bibfnamefont{I.}~\bibnamefont{Kanter}},
  \bibinfo{author}{\bibfnamefont{E.}~\bibnamefont{Kopelowitz}},
  \bibinfo{author}{\bibfnamefont{R.}~\bibnamefont{Vardi}},
  \bibinfo{author}{\bibfnamefont{M.}~\bibnamefont{Zigzag}},
  \bibinfo{author}{\bibfnamefont{W.}~\bibnamefont{Kinzel}},
  \bibinfo{author}{\bibfnamefont{M.}~\bibnamefont{Abeles}}, \bibnamefont{and}
  \bibinfo{author}{\bibfnamefont{D.}~\bibnamefont{Cohen}},
  \bibinfo{journal}{EPL (Europhysics Letters)} \textbf{\bibinfo{volume}{93}},
  \bibinfo{pages}{66001} (\bibinfo{year}{2011}{\natexlab{b}}).

\bibitem[{\citenamefont{Zigzag et~al.}(2010)\citenamefont{Zigzag, Butkovski,
  Englert, Kinzel, and Kanter}}]{Zigzag2010}
\bibinfo{author}{\bibfnamefont{M.}~\bibnamefont{Zigzag}},
  \bibinfo{author}{\bibfnamefont{M.}~\bibnamefont{Butkovski}},
  \bibinfo{author}{\bibfnamefont{A.}~\bibnamefont{Englert}},
  \bibinfo{author}{\bibfnamefont{W.}~\bibnamefont{Kinzel}}, \bibnamefont{and}
  \bibinfo{author}{\bibfnamefont{I.}~\bibnamefont{Kanter}},
  \bibinfo{journal}{Phys. Rev. E} \textbf{\bibinfo{volume}{81}},
  \bibinfo{pages}{036215} (\bibinfo{year}{2010}).

\bibitem[{\citenamefont{Englert et~al.}(2010)\citenamefont{Englert, Kinzel,
  Aviad, Butkovski, Reidler, Zigzag, Kanter, and Rosenbluh}}]{Rosenbluh2010}
\bibinfo{author}{\bibfnamefont{A.}~\bibnamefont{Englert}},
  \bibinfo{author}{\bibfnamefont{W.}~\bibnamefont{Kinzel}},
  \bibinfo{author}{\bibfnamefont{Y.}~\bibnamefont{Aviad}},
  \bibinfo{author}{\bibfnamefont{M.}~\bibnamefont{Butkovski}},
  \bibinfo{author}{\bibfnamefont{I.}~\bibnamefont{Reidler}},
  \bibinfo{author}{\bibfnamefont{M.}~\bibnamefont{Zigzag}},
  \bibinfo{author}{\bibfnamefont{I.}~\bibnamefont{Kanter}}, \bibnamefont{and}
  \bibinfo{author}{\bibfnamefont{M.}~\bibnamefont{Rosenbluh}},
  \bibinfo{journal}{Phys. Rev. Lett.} \textbf{\bibinfo{volume}{104}},
  \bibinfo{pages}{114102} (\bibinfo{year}{2010}).

\bibitem[{\citenamefont{Nixon et~al.}(2012)\citenamefont{Nixon, Friedman, Ronen,
  Friesem, Davidson, and Kanter}}]{Nixon2012}
\bibinfo{author}{\bibfnamefont{M.}~\bibnamefont{Nixon}},
  \bibinfo{author}{\bibfnamefont{M.}~\bibnamefont{Friedman}},
  \bibinfo{author}{\bibfnamefont{E.}~\bibnamefont{Ronen}},
  \bibinfo{author}{\bibfnamefont{A.~A.} \bibnamefont{Friesem}},
  \bibinfo{author}{\bibfnamefont{N.}~\bibnamefont{Davidson}}, \bibnamefont{and}
  \bibinfo{author}{\bibfnamefont{I.}~\bibnamefont{Kanter}},
  \bibinfo{journal}{Phys. Rev. Lett.} \textbf{\bibinfo{volume}{108}},
  \bibinfo{pages}{214101} (\bibinfo{year}{2012}).

\bibitem[{\citenamefont{Pecora and Carroll}(1998)}]{Pecora1998}
\bibinfo{author}{\bibfnamefont{L.~M.} \bibnamefont{Pecora}} \bibnamefont{and}
  \bibinfo{author}{\bibfnamefont{T.~L.} \bibnamefont{Carroll}},
  \bibinfo{journal}{Phys. Rev. Lett.} \textbf{\bibinfo{volume}{80}},
  \bibinfo{pages}{2109} (\bibinfo{year}{1998}).

\bibitem[{\citenamefont{Kinzel et~al.}(2009)\citenamefont{Kinzel, Englert,
  Reents, Zigzag, and Kanter}}]{Kinzel2009}
\bibinfo{author}{\bibfnamefont{W.}~\bibnamefont{Kinzel}},
  \bibinfo{author}{\bibfnamefont{A.}~\bibnamefont{Englert}},
  \bibinfo{author}{\bibfnamefont{G.}~\bibnamefont{Reents}},
  \bibinfo{author}{\bibfnamefont{M.}~\bibnamefont{Zigzag}}, \bibnamefont{and}
  \bibinfo{author}{\bibfnamefont{I.}~\bibnamefont{Kanter}},
  \bibinfo{journal}{Phys. Rev. E} \textbf{\bibinfo{volume}{79}},
  \bibinfo{pages}{056207} (\bibinfo{year}{2009}).

\bibitem[{\citenamefont{Englert et~al.}(2011)\citenamefont{Englert,
  Heiligenthal, Kinzel, and Kanter}}]{EnglertKanter2011}
\bibinfo{author}{\bibfnamefont{A.}~\bibnamefont{Englert}},
  \bibinfo{author}{\bibfnamefont{S.}~\bibnamefont{Heiligenthal}},
  \bibinfo{author}{\bibfnamefont{W.}~\bibnamefont{Kinzel}}, \bibnamefont{and}
  \bibinfo{author}{\bibfnamefont{I.}~\bibnamefont{Kanter}},
  \bibinfo{journal}{Phys. Rev. E} \textbf{\bibinfo{volume}{83}},
  \bibinfo{pages}{046222} (\bibinfo{year}{2011}).

\bibitem[{\citenamefont{J\"ungling et~al.}(2015)\citenamefont{J\"ungling,
  D'Huys, and Kinzel}}]{Jungling2015}
\bibinfo{author}{\bibfnamefont{T.}~\bibnamefont{J\"ungling}},
  \bibinfo{author}{\bibfnamefont{O.}~\bibnamefont{D'Huys}}, \bibnamefont{and}
  \bibinfo{author}{\bibfnamefont{W.}~\bibnamefont{Kinzel}},
  \bibinfo{journal}{Phys. Rev. E} \textbf{\bibinfo{volume}{91}},
  \bibinfo{pages}{062918} (\bibinfo{year}{2015}).

\bibitem[{\citenamefont{Lichtner et~al.}(2011)\citenamefont{Lichtner, Wolfrum,
  and Yanchuk}}]{Yanchuk2009}
\bibinfo{author}{\bibfnamefont{M.}~\bibnamefont{Lichtner}},
  \bibinfo{author}{\bibfnamefont{M.}~\bibnamefont{Wolfrum}}, \bibnamefont{and}
  \bibinfo{author}{\bibfnamefont{S.}~\bibnamefont{Yanchuk}},
  \bibinfo{journal}{SIAM J. Math. Anal.} \textbf{\bibinfo{volume}{43}},
  \bibinfo{pages}{788} (\bibinfo{year}{2011}).

\bibitem[{\citenamefont{D'Huys et~al.}(2012)\citenamefont{D'Huys, Fischer,
  Danckaert, and Vicente}}]{DHuys2012}
\bibinfo{author}{\bibfnamefont{O.}~\bibnamefont{D'Huys}},
  \bibinfo{author}{\bibfnamefont{I.}~\bibnamefont{Fischer}},
  \bibinfo{author}{\bibfnamefont{J.}~\bibnamefont{Danckaert}},
  \bibnamefont{and} \bibinfo{author}{\bibfnamefont{R.}~\bibnamefont{Vicente}},
  \bibinfo{journal}{Phys. Rev. E} \textbf{\bibinfo{volume}{85}},
  \bibinfo{pages}{056209} (\bibinfo{year}{2012}).

\end{thebibliography}

\end{document}